\documentclass[unsortedaddress,preprint,show pacs]{revtex4}
\usepackage{graphicx}
\usepackage{dcolumn}
\usepackage{amsmath}
\usepackage{amssymb}
\usepackage{amsfonts}
\usepackage{bm}
\usepackage{color}
\newcommand{\be}{\begin{equation}}
\newcommand{\ee}{\end{equation}}
\newcommand{\ba}{\begin{eqnarray}}
\newcommand{\ea}{\end{eqnarray}}
\begin{document}
\title{Cluster phases of penetrable rods on a line}
\author{Santi Prestipino\footnote{Email: {\tt sprestipino@unime.it}}}
\affiliation{Universit\`a degli Studi di Messina, Dipartimento di Fisica e di Scienze della Terra, Contrada Papardo, I-98166 Messina, Italy\\and\\CNR-IPCF, Viale F. Stagno d'Alcontres 37, I-98158 Messina, Italy}
\date{\today}
\begin{abstract}
Phase transitions are uncommon among homogenous one-dimensional fluids of classical particles owing to a general non-existence result due to van Hove. A way to circumvent van Hove's theorem is to consider an interparticle potential that is finite everywhere. Of this type is the generalized exponential model of index 4 (GEM4 potential), a model interaction which in three dimensions provides an accurate description of the effective pair repulsion between dissolved soft macromolecules (e.g., flexible dendrimers). Using specialized free-energy methods, I reconstruct the equilibrium phase diagram of the one-dimensional GEM4 system, showing that, apart from the usual fluid phase at low densities, it consists of an endless sequence of {\em cluster fluid phases} of increasing pressure, having a sharp crystal appearance for low temperatures. The coexistence line between successive phases in the sequence invariably terminates at a critical point. Focussing on the first of such transitions, I show that the growth of the 2-cluster phase from the metastable ordinary fluid is extremely slow, even for large supersaturations. Finally, I clarify the apparent paradox of the observation of an activation barrier to nucleation in a system where, due to the dimensionality of the hosting space, the critical radius is expected to vanish.
\end{abstract}
\maketitle

\section{Introduction}
It sometimes happens in physics that the study of one-dimensional (1D) model systems, involving much simpler mathematics than required in three dimensions, helps to shed light on the properties of ordinary three-dimensional matter. Two examples from one-particle quantum mechanics are the Kronig-Penney model of electron energy bands in a crystal and the tunneling through a 1D potential barrier that classically could not be surmounted. Statistical mechanics makes no exception, as illustrated, for instance, by the 1D model devised by Kac and coworkers~\cite{Kac} in order to derive the van der Waals theory of liquid-vapour transition from statistical mechanics; or the continuous model invented by Burkhardt for the thermal depinning of an interface bound to a 1D attractive substrate~\cite{Burkhardt}. More often, however, the 1D behavior is unlike the three-dimensional one, and in this case it is worth unveiling the mechanism by which space dimensionality makes the difference. Examples of this sort are numerous, see e.g. the 1D breakdown of the Fermi-liquid quasi-particle picture of electron transport or the general critical-point behavior which, as well known, is heavily affected by the dimensionality of the hosting space.

The situation is apparently clear for the statistical mechanics of homogeneous classical fluids with short-range interactions between the particles, thanks to an old result by van Hove which would rule out the possibility of phase transitions in one dimension~\cite{vanHove}. However, as carefully explained by Cuesta and Sanchez~\cite{Cuesta}, what is generally failed to mention in this context is a crucial assumption behind van Hove's derivation, i.e., that particles have a hard core. This opens the possibility of observing phase transitions in certain peculiar 1D systems where this constraint is violated, like for instance in systems of self-avoiding polymers flowing in a narrow channel of size comparable with the gyration radius. In this case, a coarse-grained description may be adopted where the polymers are viewed as point particles interacting through an {\em effective} potential which stays finite at the origin, implying that particles are in principle free to exchange position with each other within the line. A 1D system of this sort has recently been investigated by simulation, see Ref.\,\cite{Speranza}, but here the outcome was that no clear-cut phase transition exists in spite of a rich anomalous thermodynamic behavior. In the latter paper the particles were taken to interact via a simple Gaussian repulsion, which is sufficient to induce crystallization on cooling in both two~\cite{Prestipino1} and three dimensions~\cite{Stillinger,Prestipino2}. Here I show that, by only a slight modification of the interaction law, that is by changing from 2 to 4 the power-law exponent of the interparticle distance in the exponential argument, one succeeds in materializing several distinct {\em fluid} phases (in fact, an infinite number) in a 1D system of repulsive particles as a function of pressure. Note that {\em any} exponent strictly larger than 2 would obtain the same effect.

The $\exp(-r^4)$ model, also known as the {\em generalized exponential model} of index 4 (GEM4), was extensively studied in three dimensions~\cite{Likos1,Mladek1,Mladek2,Zhang1} where it accurately describes the effective repulsion between flexible dendrimers in a solution. As anticipated in Ref.\,\cite{Likos2}, a peculiarity of this system is to undergo {\em clustering} at high pressure, to be intended here as the formation of stable crystals with a mean number $n_c$ of particles per lattice site larger than one ($n_c$ is integer or close to an integer only for very low temperatures). These so-called {\em cluster crystals} may show a variable underlying lattice structure, either fcc or bcc. Upon compression, a cluster crystal initially resists by progressively reducing its lattice constant, until it becomes favorable for the system to eliminate lattice sites and thus transform into a different cluster phase with a higher site occupancy. At low temperature this process involves a first-order phase transition, which becomes broadened for higher temperatures (see e.g. Ref.~\cite{Wilding}).

The main feature distinguishing a cluster crystal from an ordinary crystal is the status enjoyed by the mean site occupancy $n_c$, which is not a fixed parameter but one that undergoes thermal relaxation like any other unconstrained variable. This is evident close to a phase-transition locus, where we observe a sizeable deviation of $n_c$ from an integer value, the more so the higher the system temperature. The new variable $n_c$ has a conjugate thermodynamic variable associated with it, $\mu_c$, a sort of chemical potential, which spontaneously adjusts to zero in equilibrium~\cite{Swope,Mladek1}. This fact considerably complicates the numerical determination of the free energy with respect to ordinary crystals, since a further minimization of the free energy as a function of $n_c$ has to be accomplished~\cite{Mladek1,Zhang1}.

In this paper I carry out the study of the GEM4 system in one dimension. Even though a genuine 1D crystal cannot exist at any non-zero temperature, since it would be unstable against thermal fluctuations according to Peierls' argument~\cite{Peierls}, nothing excludes the possibility of distinct fluid phases. After providing a proof of the existence of stable cluster crystals at zero temperature, I shall use Monte Carlo simulation to draw the complete 1D GEM4 phase diagram, using the method introduced in Ref.\,\cite{Mladek1} for free-energy computations. We will see that a distinct {\em cluster fluid phase} emerges out of each stable zero-temperature cluster-crystal state. In addition, a further standard (i.e., non-cluster) fluid phase is stable for low pressures. Depending on the temperature, this low-density phase resembles more either a crystal or an ordinary fluid. One major concern will be to describe how this phase dynamically transforms into the 2-cluster phase upon compression.

The outline of the paper is the following. After recalling in Sect.\,II the definition of the model and the methods employed to study it in detail, I show and discuss the simulation results in Sect.\,III. Some concluding remarks are presented in Sect.\,IV.

\section{Model and method}
\setcounter{equation}{0}
\renewcommand{\theequation}{2.\arabic{equation}}

The model potential considered here is the same as studied in \cite{Zhang1}, $u(r)=\epsilon\,\exp\{-(r/\sigma)^4\}$, where $\epsilon>0$ and $\sigma$ are arbitrary energy and length units, respectively. In one dimension, the GEM4 system can appropriately be described as a fluid of penetrable rods of diameter $\sigma$, which can fully overlap with a finite energy penalty ($\epsilon$, rather than infinity as in the case of hard rods). Since the particle ordering along the line is not preserved by the dynamics, the 1D GEM4 has no simple exact solution and it is thus necessary to study its collective behavior by numerical simulation. The Fourier transform of $u(r)$, which in 1D reads
\be
\tilde{u}(q)=2\int_0^\infty{\rm d}x\,u(x)\cos(qx)\,,
\label{2-1}
\ee
takes values of both signs; hence, in conformity with the criterion introduced in Ref.\,\cite{Likos2}, the system would support stable cluster phases at high pressure. This will shortly be confirmed by exact zero-temperature ($T=0$) calculations.

Preliminary to the simulation study of the system is the work of identifying the relevant phases, which can be done through the exact determination of the chemical potential $\mu$ as a function of the pressure $P$ at $T=0$. This was accomplished through exact total-energy calculations for a number of candidate {\em states} (regular lattice structures), as described in the following. For any given structure and $P$ value, I computed the minimum of $E+PV$ (where $E$ is energy and $V$ is volume) over a set of variables comprising the density $\rho=N/V$ and, possibly, also a number of internal parameters (see, e.g., Ref.\,\cite{Prestipino3}; more details in appendix A). I find that, on increasing the pressure, the lowest chemical potential is sequentially provided by each $n$-cluster crystal of perfectly overlapping particles ($n=1,2,\ldots$; see Fig.\,1, left panel); the transition from one state to the other always occurs with a jump in the density, signaling first-order transition behavior. For low enough pressures, the ordinary ($n_c=1$ or ``1'') crystal has the minimum $\mu$ by far. The ``1'' crystal is eventually superseded at $P=0.834$ (reduced units) by the 2-cluster crystal ($n_c=2$ or ``2'' crystal), which represents the most stable phase up to $P=2.465$; and so on (see Table 1).

We might reasonably expect that the sequence of stable phases remains unchanged for sufficiently small non-zero temperatures, would not it be for the fact that Peierls' argument rigorously excludes {\em perfect} crystal order for any $T>0$ (though in practical terms this is a mainly technical issue only relevant for very large $N$ values). A possibility is, as we shall see later, that cluster phases indeed survive for $T>0$ but in the form of {\em fluid} phases, whose structure may nonetheless easily be confused with that of a cluster crystal if temperature is not too high. However, what about the existence of other phases at $T>0$, different from cluster fluid phases with integer $n_c$? This is a possible occurrence only if at $T=0$ some other phase is close in chemical potential to the cluster crystals with integer $n_c$. Indeed, this is exactly what happens near each of the transition pressures listed in Table 1. As a matter of example, I have plotted in the right panels of Fig.\,1 the chemical potential of a number of periodic structures with single particles and pairs in various combinations (see appendix A for notation), in a range of pressures around $P=0.834$. It appears that the chemical potentials of all these crystals, including ``1'' and ``2'', become nearly the same for $P\simeq 0.834$. What is amazing is that crystal states with vastly different $n_c$ values, all in the range from 1 to 2, are nearly degenerate near $P=0.834$, suggesting a non-trivial transition scenario for $T>0$ characterized by a very slow relaxation dynamics. I shall come back to this issue later, at the end of Sect.\,III, where the kinetic pathways of the phase transformation from ``1'' to ``2'' will directly be probed by simulation.

%
%
\begin{table}
\caption{Zero-temperature phases of the 1D GEM4 system (only data the first four phases are listed). For each pressure range in column 1, the thermodynamically stable phase is indicated in column 3, together with the respective values of the number density $\rho$ (column 2).}
\begin{tabular*}{\columnwidth}[c]{@{\extracolsep{\fill}}|c|c|c|}
\hline
$P$ range ($\epsilon/\sigma$) & $\rho$ range ($\sigma^{-1}$) & stable phase \\
\hline\hline
0-0.833 & 0-0.8268 & ``1'' \\
\hline
0.834-2.465 & 1.4149-1.5778 & ``2'' \\
\hline
2.466-4.913 & 2.1701-2.3295 & ``3'' \\
\hline
4.914-8.178 & 2.9228-3.0812 & ``4'' \\
\hline
\end{tabular*}
\end{table}

The phase diagram of the 1D GEM4 system was mainly investigated by Monte Carlo (MC) simulation in the $NPT$ (isothermal-isobaric) ensemble, with not less than $N=400$ particles and periodic boundary conditions (in a few cases I checked that finite-size effects are negligible). Besides the standard local particle moves, in order to speed up the MC sampling an average 50\% of the displacements were directed towards randomly chosen positions in the simulation box. Typically, for each $(T,P)$ state as many as $2\times 10^6$ cycles or sweeps (i.e., moves per particle) were generated at equilibrium, which proved to be many enough to obtain accurate statistical averages for the volume and the energy per particle (at least, far from critical points). Much longer production runs of 5 to 50 million sweeps each were performed at closely-separated state points on a few isotherms in order to study the transformation from ``1'' to ``2'' and {\it vice versa} (see Sect.\,III). A run of 5 million sweeps at $P=0.5$ and $T=0.1$ turned out to be sufficient to accurately compute the chemical potential of the ``1'' phase by Widom's particle-insertion method~\cite{Widom}. The location of each phase transition was determined through thermodynamic integration of chemical-potential derivatives along isobaric and isothermal paths connecting the system of interest to a reference system whose free energy is already known (see, e.g., Ref.~\cite{Saija1}). While the reference state for the ``1'' phase was a dilute fluid, deep inside each cluster region a low-temperature cluster fluid with integer $n_c$ was taken as the starting point of a MC trajectory. In any such state, the Helmholtz free energy was numerically computed by a variant~\cite{Mladek1} of the Einstein-crystal method which is briefly described below.

In the original Frenkel-Ladd method~\cite{Frenkel,Polson}, the free energy of the system of interest is built up starting from the known free energy of a system of independent harmonic oscillators. This is realized by introducing a linear morphing $U_\lambda$ ($0\le\lambda\le1$) between the two potential energies, and computing for each intermediate step $\lambda$ the average energy difference $\langle\Delta U\rangle_\lambda$ between the two systems in a $NVT$ simulation. To prevent $\langle\Delta U\rangle_\lambda$ from diverging at $\lambda=1$ (corresponding to $U_\lambda=U$, the actual potential energy), all simulations are performed under the constraint of a fixed center of mass and the bias thus introduced is corrected at the end of the calculation. In simulations of a crystal with multiply-occupied sites and itinerant particles ceaselessly hopping from site to site, a more appropriate choice of reference system is an ideal gas of particles diffusing in a landscape of well-separated potential barriers centered on the lattice sites~\cite{Mladek1}, and there is no longer need to constrain the center of mass. Note that the same modified Frenkel-Ladd method can also be employed to compute the chemical potential of a cluster fluid, the only difference being that the morphing formula now relates the free energies of two distinct fluid systems in the same thermodynamic state. As conveniently argued in Ref.\,\cite{Mladek2}, in order to improve sampling efficiency for $\lambda=1$ on average one MC move out of $N$ was attempted to shift the system center of mass to a totally random position. For other values of $\lambda$, the trial moves were of the same type as considered for $NPT$ runs. Particular care was paid in the extraction of $F$ from $\langle\Delta U\rangle_\lambda$, since the latter quantity exhibits a strong dependence on $\lambda$ near 0 and 1 (see the typical low-$T$ profile of this function in Fig.\,2).

For non-integer $n_c$ values, a delicate issue is the choice of the initial configuration of a $NPT$ or $NVT$ simulation. In a few cases I found that an unfortunate arrangement of the particles drives the GEM4 system to a long-lived out-of-equilibrium configuration characterized by an unphysical density wave. For $1<n_c<2$, I eventually realized that the right choice is to randomly mix, in the correct proportions, single particles and fully overlapping pairs. On the contrary, the choice of dropping $N$ particles at random on $M$ sites (with $n_c=N/M$) sometimes led to inefficient sampling.

Clearly, by the above outlined method one just computes the Helmholtz free energy $F(n_c)$ of the system for a preset $n_c$ value {\em and} a given state point. In principle, at each $(P,T)$ point one should repeat the free-energy calculation for several $n_c$'s and eventually pick the $n_c$ value with the minimum associated $\mu$. This is a rather daunting task, which however can be avoided if thermodynamic integration is employed judiciously: it would actually be sufficient to compute $\mu$ at one reference state for several values of $n_c$ and then generate a separate (either isobaric or isothermal) chain of Monte Carlo runs for each $n_c$. The actual system chemical-potential curve along the given $(P,T)$ path is the lower envelope of the various $\mu(n_c)$ curves. This is a correct procedure which, however, is doomed to fail if $n_c$ abruptly changes along the path. Even in cases where $n_c$ is perfectly conserved, one should nonetheless check -- by, e.g., visual inspection on a random basis -- that the spatial mixing of 1's and 2's is really random at each state point along the path.

A further tool employed to investigate the system phase diagram at high temperature was density-functional theory (DFT) in the mean-field (MF) approximation~\cite{Evans,Likos1}, which is thoroughly presented in the appendix B. Here I just observe that the efficacy of such a theory (indeed quite accurate in three dimensions) typically diminishes on reducing the dimensionality of the system, but it will anyway be interesting to see how the 1D theory performs compared to three dimensions.

\section{Results}
\setcounter{equation}{0}
\renewcommand{\theequation}{3.\arabic{equation}}

\subsection{The equilibrium phase diagram}
The exact calculations of Sect.\,II only apply for $T=0$. To get some insight into the behavior of the 1D GEM4 at $T>0$, I carried out a series of concatenated $NPT$ runs along a number of isothermal paths, initially in large steps of $\Delta P=0.1$, starting from a disordered configuration at low pressure and from an ordered cluster-crystal configuration at high pressure. I employed 2 million MC sweeps to compute equilibrium averages at each state point, while another 2 million sweeps were used to equilibrate the system from the last configuration generated at the preceding state in the chain. In Fig.\,3 the system density $\rho$ is plotted as a function of the reduced pressure, for a few $T$ values in the range from $0.01$ to $0.07$. As neatly shown by these graphs, upon compression the system crosses a series of phase-coexistence lines at nearly the same transition pressures as found at $T=0$, with more and more extended hysteresis loops as $P$ grows. On increasing the temperature, the transitions become rounded one after another in the same order, suggesting that the coexistence loci all terminate with a critical point ($T_c\simeq 0.02$ for the ``1'' to ``2'' transition, $T_c\simeq 0.05$ for the ``2'' to ``3'' transition, and so on), as if they were standard liquid-vapor loci. In 3D, the critical point of each isostructural $n_c\leftrightarrow(n_c+1)$ transition belongs to the Ising universality class~\cite{Zhang2}. The eventual coalescence of curves with a different attached $n_c$ index proves that the mean site occupancy is not conserved along a chain of runs but rather changes with pressure, at least across a phase transition. We will see that the flexibility of $n_c$ is even more pronounced than the above evidence indicates. As a matter of fact, $n_c$ promptly reacts to {\em any} change of thermodynamic control parameters, a circumstance that, rather than being a problem, actually considerably simplifies the numerical determination of the phase boundaries for the present 1D system.

Next I focussed on the transition regions and accordingly reduced the pressure step $\Delta P$ while leaving the rest of the simulation schedule unchanged. By combining the free-energy values obtained at selected reference states with the density and energy data collected along a few isothermal and isobaric paths on the $(P,T)$ plane, I eventually obtained the tentative phase diagram schematized in Fig.\,4 (lower panels). Overall, this behavior conforms with the picture already emerged from Fig.\,3. In the upper panels of Fig.\,4, I have reported the outcome of the MF-DFT theory described in appendix B, which predicts a transition from fluid to cluster crystal (our DFT calculations are blind to Peierls' argument).

A closer look at Fig.\,4 reveals a potential matching problem between the high- and low-temperature behaviors: specifically, what is the true nature of the transition predicted by DFT and, moreover, where is the origin of the transition line derived from it? To find an answer I carried out a series of simulations at $T=0.5$, across the alleged phase transition point at $P=7.85$. Assuming a stable cluster phase around this pressure, I first tried to establish the equilibrium value of $n_c$ at $\rho=3.5$ (corresponding to $P\simeq 11.91$). To this aim I computed $\langle\Delta U\rangle_\lambda$ for $n_c=1,2,\ldots 6$. Notwithstanding these curves are quite different from one another (see Fig.\,5), they all sum up to practically the same $\mu$ (i.e., the same within $\approx 0.004$ in the whole $P$ range between 5 and 12), suggesting that the system is actually not a cluster phase at these pressures but rather an ordinary fluid. A close inspection into the typical system configuration confirmed this: particles are nearly homogeneously dispersed along the line, with no blobs or lumps at regular intervals (i.e., in the radial distribution function (RDF) profile the $r=0$ peak is not neatly disjoint from the next peak). Hence, the slight diversity of $\mu$ between the various $n_c$'s would just be due to the use of distinct $N$'s for the different $n_c$'s. In Fig.\,6 I have reported the energy and density data computed for the various $n_c$'s, showing no clear sign of a phase transition (at least up to $P=25$). In the same Fig.\,6 I have also reported the system density derived from MF-DFT. Apart from the cusp at $P=7.85$, which alludes to a nonexistent phase transition, the DFT density is not bad at all if compared with the MC results. Clearly, for much larger pressures we might hit the coexistence line for some high-index, $n_c\leftrightarrow(n_c+1)$ transition, and this would clearly entail a jump discontinuity for $\rho$, which anyway could be hard to locate considering hysteresis.

In conclusion, the phase diagram of the 1D GEM4 system is as represented in the lower panels of Fig.\,4, with an infinite sequence of finite-length transition loci separating {\em fluid} phases of increasing pressure; therefore, it will be possible to go from one phase to another without ever crossing a first-order line. Fig.\,7 shows the RDF of a $N=400$ system at $T=0.01$ and $P=1.6$. This graph shows that at low temperature a 2-cluster fluid retains a sharp spatial periodicity up to relatively large distances. Therefore, if temperature is not too high it would be hard to distinguish a cluster-fluid system from a cluster crystal only by looking at its local structure.

\subsection{A closer inspection into the critical region}
Among the transition points reported in the lower panels of Fig.\,4, those requiring a follow-up investigation are the nearest to the critical points, since the asymmetry observed near the top of the binodal line is a dubious feature which might be the outcome of an insufficient equilibration. Therefore I repeated the simulations at $T=0.016$ (``1''-``2'' transition) and $T=0.05$ (``2''-``3'' transition), allowing for longer equilibration and production runs.

The behavior of the 1D GEM4 system at $T=0.016$ is best considered in comparison with that for $T=0.02$ (Fig.\,8 top). At the latter temperature, no hysteresis is found: the energy and density data for a system initially prepared in the ``1'' state do exactly overlap with those found by gradually decompressing the system from a perfect ``2'' state. The energy and density data do not even change when the initial cluster-crystal state corresponded to $n_c=1.9$ or 1.8, suggesting that the system is capable to fast reabsorb a few randomly dispersed single-particle defects, thus adapting their number to the thermodynamic conditions. In these $T=0.02$ simulations the minimum number of MC sweeps per point was $2\times 10^6$, as before, plus an equal number of discarded cycles used to equilibrate the system from the preceding state. The outcome for $T=0.016$ using the same number of sweeps was totally different and more consistent with a first-order transition. However, the true point is that the latter scenario changes qualitatively when the simulation schedule is different: I found that the compression and decompression trajectories followed by the system at $T=0.016$ are a function of the number of sweeps generated per point, see Fig.\,8 center. When this number eventually overcomes $5\times 10^7$, the $1\leftrightarrow 2$ transformation becomes perfectly smooth, implying that the critical temperature for the $1\leftrightarrow 2$ transition is in fact smaller than 0.016.

Another example is $T=0.01$ (see Fig.\,8 bottom): now the behavior clearly points to the existence of a density jump at the transition, but the size of the hysteresis loop (i.e., the ``diameter'' of the spinodal region) reduces progressively as the number of sweeps per point grows larger. Admittedly, the system equilibration is extremely slow at low temperature, at least near above and across a phase-transition line, in fact so much slow that it would be difficult to decide numerically where the critical temperature of the $1\leftrightarrow 2$ transition is exactly located (let alone the value of the critical exponents). Summarizing, 1) the critical temperature $T_c$ for the ``1''-``2'' transition lies somewhere between 0.01 and 0.016; 2) the diameter of the binodal region stays roughly constant up to about $T=0.01$, but it would then shrink fast upon heating from 0.01 to $T_c$. To find some traces of the criticality at $T_c$ I have plotted in Fig.\,9 the isobaric specific heat $c_P$ and the isothermal compressibility $K_T$ as a function of the pressure along the isotherms at $T=0.016$ and $T=0.02$. Notwithstanding the large statistical errors, the reported behavior is clearly suggestive of the crossing of a Widom line.

The smoothness of the system number density slightly above $T_c$ suggests a similarly smooth evolution for the mean site occupancy $n_c$, which is taken to be the average ratio of the particle number $N$ to the number of ``sites'' $N_s$. Rather than estimating $N_s$ for each system configuration, we may look at the RDF peak centered at zero, which is present whenever the number of paired particles in the system is non-zero on average. When $1\le n_c\le 2$, the number of particle pairs is $N-N_s$, hence $N-2(N-N_s)=2N_s-N$ is the number of unpaired or isolated particles. If $x_0$ denotes the width of the first RDF peak, the average number $N(r<x_0)$ of particles whose distance from a reference particle at the origin is less than $x_0$ is:
\be
2\rho\int_0^{x_0}{\rm d}x\,g(x)=N(r<x_0)=\frac{2N_s-N}{N}\times 0+\left(1-\frac{2N_s-N}{N}\right)\times 1=2\left(1-\frac{N_s}{N}\right)\approx 2-\frac{2}{n_c}\,,
\label{3-1}
\ee
assuming that the distribution of $N_s$ is sharply peaked. Using the above formula, I estimated the values of $n_c$ along the isothermal paths of Fig.\,8 (top and center only), finding a behavior very similar to that of the density, see Fig.\,10.

Next, I extended the above analysis to the transition between ``2'' and ``3'' along the $T=0.05$ path. Again, when the simulation runs take a longer time, the transformation that previously looked first-order now slowly glides into a continuous crossover, see Fig.\,11. Here, the compression trajectory for the most accurate sampling does not exactly overlap with the decompression trajectory, but this could simply be related to the difficulty in equilibrating the system structure at these high pressures, which would require a much smaller compression/decompression rate. The mean site occupancy is reported in the bottom panel of Fig.\,11; again, its behavior closely follows that of the density. Note that the relation of $n_c$ to $N(r<x_0)$ for $2\le n_c\le 3$ is no longer given by Eq.\,(\ref{3-1}), since a different argument now applies. Considering that the number of paired particles in a configuration with $N_s$ sites is $N-3(N-2N_s)=6N_s-2N$, it follows that
\be
N(r<x_0)=\frac{6N_s-2N}{N}\times 1+\left(1-\frac{6N_s-2N}{N}\right)\times 2=4-\frac{6N_s}{N}\approx 4-\frac{6}{n_c}\,.
\label{3-2}
\ee

\subsection{The growth of the 2-cluster phase from the fluid}
Finally, I have considered the issue of nucleation and growth in the formation of the 2-cluster phase from the metastable ``1'' phase. In 3D, a first-order phase transformation (e.g., the freezing of a metastable liquid) typically occurs via the nucleation and subsequent growth of a sufficiently large embryo of the stable phase (solid nucleus)~\cite{Kashchiev}. Indeed, some free energy must be paid for the creation of the interface between a {\em cluster} of solid-like particles (i.e., a bunch of particles, not to be confused with a single blob of nearly overlapping particles in a cluster phase) and its liquid environment, which implies that the system has to overcome an activation barrier in order that the solid can grow. The cluster free energy $\Delta G$ (i.e., the reversible work of cluster formation) is the difference between the free-energy cost of a liquid with and without a solid cluster inside. This roughly amounts to
\be
\Delta G=-|\Delta\mu|N_s+g_0N_s^{2/3}\,,
\label{3-3}
\ee
where $N_s$ is the number of solid particles, $\Delta\mu=\mu_s-\mu_l<0$ is the chemical-potential difference between the phases and $g_0$ is a ${\cal O}(k_BT)$ energy (``surface tension''). The function (\ref{3-3}) reaches a maximum at a non-zero (critical) value of $N_s$, $N_s^*\propto g_0^3/\Delta\mu^2$, implying that the lifetime of the metastable liquid is very long for low supercoolings.

Notwithstanding Eq.\,(\ref{3-3}) is in many cases inaccurate (see, e.g., Ref.\,\cite{Prestipino4}), it captures most of the nucleation phenomenon. The argument leading to (\ref{3-3}) is easily adapted to other spatial dimensionalities, simply by changing the exponent $2/3$ into $1/2$ (2D) or $0$ (1D). In particular, the expected critical cluster size in 1D would be $N_s^*=0$, meaning that there would be no time delay for the growth of the stable phase from within the parent phase after the quench. By the way, note that for any dimensionality it is far more convenient (at least sufficiently close to coexistence) that the solid grows from a single nucleus rather than from many small crystallites -- this follows from the fact that if many, ${\cal O}(N_s)$ independent clusters of size $n_{cl}$ form out of a number $N_s$ of solid particles, the surface term in Eq.\,(\ref{3-3}) has to be replaced with a larger $g_0N_s/n_{cl}^{1/3}$ term plus a positive ${\cal O}(N_s\ln N_s)$ ideal-gas free energy term~\cite{tenWolde1}.

The absence of an activation barrier to nucleation in 1D should not anyway be intended as an argument for the promptness of the transformation of the parent phase into the stable phase. This process will be fast only provided also the {\em growth} of the stable phase meets no hindrance. With specific regard to the formation of the 2-cluster phase from the 1D GEM4 fluid, the existence of many configurations with nearly the same $T=0$ chemical potential as ``1'' and ``2'' but widely different $n_c$ values (see Sect.\,II) is sufficient condition for a very slow growth kinetics. This has already been discussed in relation to Fig.\,8, where the way isotherms look slightly above $T_c$ is strongly dependent on the length of the simulation runs. Below $T_c$, the large extension of the hysteresis loops indicates that the complete transformation from one phase to the other also takes very long (simulation) time, in spite of the lack of a nucleation barrier to climb up.

I studied the growth dynamics of the 1D GEM4 system at two temperatures, $T=0.01$ and $T=0.016$, at pressures where the relaxation to equilibrium occurs via the gradual appearance of the ``2'' phase from within the ``1'' phase. {\em Clusters} of the ``2'' phase in a prevalently ``1'' phase were defined as patches of adjacent particle pairs, each ``pair'' representing two nearly overlapping soft rods. Within a standard $NPT$ simulation run, the statistics of the cluster size $k$ was computed as a function of MC time at regular intervals of $5\times 10^7$ sweeps, by gathering together data from the last $5\times 10^7$ sweeps, up to a total time of $3\times 10^8$ sweeps. Denoting ${\cal N}(k)$ the mean number of $k$-sized clusters in the system, for fixed $T$ and $P$ I have considered the time evolution of $G(k)=-\ln({\cal N}(k)/N)$, a quantity akin to a cluster free energy~\cite{tenWolde2,Prestipino5}, starting from the last configuration obtained in the original run and carefully distinguishing between the various compression trajectories in Fig.\,8. To have a hint on the overall system structure, I also computed as a function of the simulation time the total fraction $f_2$ of particle pairs in the system. Two cases were examined in detail, $T=0.016,P=0.89$ and $T=0.01,P=1.05$, which are illustrated in Fig.\,12 and 13, respectively. At these values of $T$ and $P$, particles are either isolated or strictly paired (the fraction of triplets is negligible) and the RDF peak centered at the origin is well separated from the next peak. We see that at $T=0.016$ the system gradually and very slowly gets rid of isolated particles until their number reduces to a small value which is compatible with the $n_c$ value reported in Fig.\,10 below. The quantity $G(k)$ evolves differently with MC time for the different compression trajectories, reaching eventually the same positive, nearly constant slope as a function of $k$. The outcome is similar at the lower temperature $T=0.01$, where again $G(k)$ is an increasing function of $k$ at all times. This may sound strange, since the critical cluster size would then be non-zero. The reason why these $G(k)$ plots are not in contradiction with the expectation of a zero critical-cluster size has partly to do with our cluster definition, which is not sufficiently flexible to envisage the possibility of defects (i.e., isolated particles) within a cluster of the ``2'' phase. Note that these defects would not cost too much, considering that the mixing of isolated particles and pairs in any proportions occurs at $T=0$ with only a marginal increase of the chemical potential. According to our cluster definition, very large clusters are strongly suppressed since it seldom occurs that defects are few and well concentrated in space (typical final configurations of the runs are illustrated in Fig.\,14). For reasons related to the shape of the cluster-size statistics in a fully uncorrelated mixture of 1' s and 2' s, also the occurrence of small-sized clusters in the system gets enhanced from our cluster notion, with the result that no safe indication on the cluster formation energy can be got from the $G(k)$ plots of Figs.\,12 and 13. A more subtle explanation has to do with the kind of configurations participating in the calculation of $G(k)$: rather than quasi-equilibrium configurations like in a metastable system before the first critical cluster forms, all the states over which averages were computed are fully non-equilibrium post-nucleation states, which are not those relevant for the calculation of the free energy of the nucleation cluster.

\section{Conclusions}

In the present study, the thermodynamic and growth behaviors of the 1D GEM4 system were investigated by MC simulation. The GEM4 potential is bounded at the origin, hence it escapes the conclusions of van Hove's theorem which prescribes the lack of phase transitions in 1D homogeneous classical fluids of hard-core particles with additional short-range interactions between them. Specialized numerical free-energy methods were employed in order to map out the equilibrium phase diagram of the system, thus revealing an endless sequence of cluster fluid phases beyond a low-density, ordinary fluid phase. First-order lines ending with a critical point separate adjacent phases in the sequence. In the present 1D system, the cluster phases have a pronounced crystal character for not too high temperatures.

That said, the simulation also revealed interesting (and partly unconventional) features of the system collective behavior near a phase-coexistence line: 1) a very slow relaxation to equilibrium slightly above $T_c$, maybe just a mark of the closeness to criticality, which however may lead to confuse a smooth crossover between the phases with a discontinuous transition if MC trajectories are not long enough; 2) below $T_c$, the decay of the metastable phase is extremely slow, even for huge supersaturations, and occurs without a nucleation barrier to overcome. The growth rate of the stable phase is very much sensitive to the structure of the initial system configuration, and even the cluster-size statistics is unorthodox, though this would just be the effect of the naive cluster definition used and, moreover, of the intrinsic non-equilibrium character of the configurations involved in the averaging process. I surmise that the kind of phenomenology here reported can in principle be detected in systems of dissolved polymers confined in narrow channels.

\section*{Acknowledgements}
Useful discussions with Domenico Gazzillo, Paolo V. Giaquinta, and Andr\'es Santos are gratefully acknowledged.

\appendix
\section{Zero-temperature calculations}
\setcounter{equation}{0}
\renewcommand{\theequation}{A.\arabic{equation}}

In this appendix, I provide some details on the calculation of the $P$-dependent chemical potential of various 1D GEM4 lattice structures at $T=0$. Strictly periodic configurations permit the exact computation of the system chemical potential, reducing it to seeking the minimum of a few-parameter analytical expression. For $n$-cluster crystals ($n=1,2,\ldots$), the chemical potential is easily found to be
\be
\mu=\min_{\rho}\left\{\frac{n-1}{2}\epsilon+n\epsilon\sum_{k=1}^\infty\exp\left[-\left(n\frac{k}{\rho}\right)^4\right]+\frac{P}{\rho}\right\}\,,
\label{a-1}
\ee
where the first term within parentheses is the on-site energy per particle while the second is related to interactions between particles placed at different sites.

For more complex lattices, the unit cell contains two or more sites and the energy calculation is more lengthy though in principle still straightforward. For simplicity, I only consider here 1D lattices interpolating between ``1'' and ``2''. For instance, in the lattice denoted ``12'', isolated particles (1's) occur regularly alternated with pairs of fully overlapping particles (2's) and the chemical potential turns out to be:
\be
\mu=\min_{\rho}\left\{\frac{1}{3}\epsilon+\frac{4}{3}\epsilon\sum_{k=0}^\infty\exp\left[-\left(\frac{3}{2\rho}+\frac{3k}{\rho}\right)^4\right]+\frac{5}{3}\epsilon\sum_{k=1}^\infty\exp\left[-\left(\frac{3k}{\rho}\right)^4\right]+\frac{P}{\rho}\right\}\,.
\label{a-2}
\ee
Note that the number density $\rho$ is still the only parameter to be optimized in the above formula.

The computational effort is higher for other simple lattices, like ``112'' (where the succession of particles along the line is $\cdots-1-1-2-1-1-2-\cdots$) or ``1112''. For, e.g., ``112'', two parameters (for example, the density and the distance $d_{11}$ between two successive 1's) should be tuned for each $P$, and the chemical potential reads:
\ba
&& \mu=\min_\rho\left\{\min_{0<d_{11}<4/\rho}\mu(\rho,d_{11})\right\}\,\,\,\,\,\,{\rm with}
\nonumber \\
&& \mu(\rho,d_{11})=\frac{1}{4}\epsilon+\frac{1}{4}\epsilon\sum_{k=0}^\infty e^{-\left(d_{11}+4k/\rho\right)^4}+\epsilon\sum_{k=0}^\infty e^{-\left(d_{12}+4k/\rho\right)^4}+\epsilon\sum_{k=0}^\infty e^{-\left(d_{11}+d_{12}+4k/\rho\right)^4}
\nonumber \\
&+&\frac{1}{4}\epsilon\sum_{k=0}^\infty e^{-\left(2d_{12}+4k/\rho\right)^4}+\frac{3}{2}\epsilon\sum_{k=1}^\infty e^{-\left(4k/\rho\right)^4}+\frac{P}{\rho}\,,
\label{a-3}
\ea
where $d_{12}=2/\rho-d_{11}/2$.

The calculation of $\mu$ for ``1122'' is one step more difficult. Now, we may assume the distances between two nearby 1's ($d_{11}$) and between two nearby 1 and 2 ($d_{12}$) as additional parameters to be optimized besides the density $\rho$. In terms of these parameters, the chemical potential is given by
\ba
&& \mu=\min_\rho\left\{\min_{0<d_{12}<3/\rho}\left\{\min_{0<d_{11}<6/\rho-2d_{12}}\mu(\rho,d_{11},d_{12})\right\}\right\}\,\,\,\,\,\,{\rm with}
\nonumber \\
&& \mu(\rho,d_{11},d_{12})=\frac{1}{3}\epsilon+\frac{1}{6}\epsilon\left\{\sum_{k=0}^\infty e^{-\left(d_{11}+6k/\rho\right)^4}+4\sum_{k=0}^\infty e^{-\left(d_{12}+6k/\rho\right)^4}+4\sum_{k=0}^\infty e^{-\left(d_{22}+6k/\rho\right)^4}\right\}
\nonumber \\
&+&\frac{1}{6}\epsilon\left\{4\sum_{k=0}^\infty e^{-\left(d_{11}+d_{12}+d_{22}+6k/\rho\right)^4}+\sum_{k=0}^\infty e^{-\left(2d_{12}+d_{22}+6k/\rho\right)^4}+4\sum_{k=0}^\infty e^{-\left(d_{11}+2d_{12}+6k/\rho\right)^4}\right\}
\nonumber \\
&+&\frac{2}{3}\epsilon\left\{\sum_{k=0}^\infty e^{-\left(d_{11}+d_{12}+6k/\rho\right)^4}+\sum_{k=0}^\infty e^{-\left(d_{12}+d_{22}+6k/\rho\right)^4}\right\}+\frac{5}{3}\epsilon\sum_{k=1}^\infty e^{-\left(6k/\rho\right)^4}+\frac{P}{\rho}\,,
\label{a-4}
\ea
where $d_{22}=6/\rho-d_{11}-2d_{12}$.

The chemical-potential data for all the considered lattices are reported in Fig.\,1 and commented at length in Sect.\,II.

\section{Density-functional theory of the GEM4}
\setcounter{equation}{0}
\renewcommand{\theequation}{B.\arabic{equation}}

I hereafter report a number of MF-DFT calculations for the GEM4, both in one and three dimensions, aimed at locating the system phase boundaries at sufficiently high temperature. I will mainly follow the presentation in Ref.\,\cite{Prestipino6}, though I anticipate that the results for the 3D system are substantially equivalent to those reported in \cite{Likos1} where a slightly different formalism has been developed -- built around the $NVT$ ensemble, rather than the $\mu VT$ ensemble employed here. I will also abstain from the further approximations that were made in Ref.\,\cite{Likos1} in an effort to obtain a closed-form expression for the free-energy functional.

The premise of any DFT approach to the freezing transition is to write an (approximate) grand-potential functional $\Omega_\mu$ of the crystal one-point density $n({\bf x})$, the minimization of which reveals the thermodynamics {\em and} structure of the periodic solid. On general grounds, $\Omega_\mu[n]$ is written (in three dimensions) as
\be
\Omega_\mu[n]=F^{\rm id}[n]+F^{\rm exc}[n]-\mu\int{\rm d}^3x\,n({\bf x})\,,
\label{b-1}
\ee
with separate, ideal (known) and excess (generally unknown) contributions to the Helmholtz free-energy functional $F[n]$. In the so-called Ramakrishnan-Yussouff theory~\cite{Ramakrishnan}, the two-point direct correlation function (DCF) of the crystal is approximated with that, $c(r;\rho)$, of the homogeneous fluid of density $\rho$, and the excess free energy thus reads:
\be
\beta F^{\rm exc}[n]=\beta F^{\rm exc}(\rho)-c_1(\rho)\int{\rm d}^3x(n({\bf x})-\rho)-\frac{1}{2}\int{\rm d}^3x\,{\rm d}^3x'\,c(|{\bf x}-{\bf x}'|;\rho)(n({\bf x})-\rho)(n({\bf x}')-\rho)\,,
\label{b-2}
\ee
where $c_1(\rho)=\ln(\rho\Lambda^3)-\beta\mu$ is the one-point DCF of the fluid, $\Lambda$ is the thermal wavelength, and $F^{\rm exc}(\rho)$ is the excess free energy of the fluid. A further simplifying, MF-like assumption is $c(r)\approx-\beta u(r)$, which gives $F^{\rm exc}$ the more compact form
\be
F^{\rm exc}[n]=\frac{1}{2}\int{\rm d}^3x\,{\rm d}^3x'\,u(|{\bf x}-{\bf x}'|)n({\bf x})n({\bf x}')\,.
\label{b-3}
\ee
This completes our hierarchy of approximations.

Now let $M$ and $N_s$ denote the number of lattice sites and the average number of solid particles, respectively (the two numbers are distinct in a cluster crystal). In terms of the solid density $\rho_s=N_s/V$ and the unit-cell volume $v_0=V/M$, the average site occupancy reads $n_c=N_s/M=\rho_sv_0$. A popular {\em ansatz} for $n({\bf x})$ is
\be
n({\bf x})=n_c\left(\frac{\alpha}{\pi}\right)^{3/2}\sum_{\bf R}e^{-\alpha({\bf x}-{\bf R})^2}=\rho_s\sum_{\bf G}e^{-G^2/(4\alpha)}e^{i{\bf G}\cdot{\bf x}}\,,
\label{b-4}
\ee
where the last equality transforms a direct-lattice sum into a reciprocal-lattice sum. Plugging Eq.\,(\ref{b-4}) into (\ref{b-1}), one eventually obtains the grand-potential difference $\Delta\Omega_\mu[n]=\Omega_\mu[n]-\Omega(\rho)$ between the crystal and the fluid as a function of three parameters ($v_0,\alpha$, and $n_c$), which has to be made minimum for each $\rho$. For the given $\mu,V,T$, the freezing transition occurs where the minimum $\Delta\Omega_\mu$ ($\equiv\Delta\Omega^*$) happens to be zero (the corresponding value of $\rho$ is the freezing density $\rho_f$). An explicit expression of $\Delta\Omega_\mu[n]$ is
\be
\frac{\beta\Delta\Omega_\mu}{V}=\frac{1}{V}\int{\rm d}^3x\,n({\bf x})\ln\frac{n({\bf x})}{\rho}-(\rho_s-\rho)+\frac{\beta}{2}(\rho_s-\rho)^2\widetilde{u}(0)+\frac{\beta}{2}\rho_s^2\sum_{{\bf G}\ne 0}e^{-G^2/(2\alpha)}\widetilde{u}({\bf G})
\label{b-5}
\ee
with
\be
\widetilde{u}(0)=4\pi\int_0^\infty{\rm d}r\,r^2u(r)\,\,\,\,\,\,{\rm and}\,\,\,\,\,\,\widetilde{u}({\bf G})=4\pi\int_0^\infty{\rm d}r\,r^2u(r)\frac{\sin Gr}{Gr}\,.
\label{b-6}
\ee
In three dimensions, the high-precision evaluation of the integral in Eq.\,(\ref{b-5}) can be made by the method explained in \cite{Wallace} (see also Ref.\,\cite{Prestipino7}).

From the general formulae
\be
c_1'(\rho)=\int{\rm d}^3x\,c(|{\bf x}-{\bf x}'|;\rho)=\widetilde{c}(0;\rho)\,\,\,\,\,\,{\rm and}\,\,\,\,\,\,\beta f^{\rm exc}(\rho)=-\frac{1}{\rho}\int_0^\rho{\rm d}\rho'(\rho-\rho')\widetilde{c}(0;\rho')\,,
\label{b-7}
\ee
it follows in the mean-field approximation that
\be
c_1(\rho)=-\beta\widetilde{u}(0)\rho\,\,\,\,\,\,{\rm and}\,\,\,\,\,\,\beta f^{\rm exc}(\rho)=\frac{\beta\widetilde{u}(0)}{2}\rho\,.
\label{b-8}
\ee
Hence, the solid grand potential can be written as
\be
\frac{\beta\Omega[n]}{V}=\frac{\beta\Delta\Omega[n]}{V}-\rho-\frac{\beta\widetilde{u}(0)}{2}\rho^2\,.
\label{b-9}
\ee
At equilibrium, the fluid and solid pressures are finally given in terms of the fluid density $\rho$ by
\be
\beta P_{\rm fluid}=-\frac{\beta\Omega(\rho)}{V}=\rho+\frac{\beta\widetilde{u}(0)}{2}\rho^2\,\,\,\,\,\,{\rm and}\,\,\,\,\,\,\beta P_{\rm solid}=\beta P_{\rm fluid}-\frac{\beta\Delta\Omega^*(\rho)}{V}\,.
\label{b-10}
\ee

I report a few results for the 3D GEM4 system in Fig.\,15 and the resulting phase diagram in Fig.\,16 (to be compared with that in Fig.\,4 of Ref.\,\cite{Likos1} and with the exact one in Ref.\,\cite{Zhang1}). On the basis of the present theory, for all temperatures down to $T=0.1$ the phase sequence upon compression is fluid-(cluster bcc)-(cluster fcc), in agreement with the exact MC results by Zhang {\em et al.}~\cite{Zhang1}. Within each cluster phase, the density $\rho_s$, the $\alpha$ value, and the mean site occupancy $n_c$ all increase almost linearly with pressure.

Adapting the above DFT theory to the 1D case is a trivial task (the Fourier transform is now defined by Eq.\,(\ref{2-1})). Clearly, since no 1D crystal would truly exist, what I am assuming here is that the {\em thermodynamics and local structure} of a 1D cluster fluid are not too dissimilar from those of a (fictitious) cluster crystal. I show some MF-DFT results for the 1D GEM4 system in Fig.\,17. Again, the theory predicts a phase transition from fluid to cluster crystal but, at variance with the 3D case, freezing now occurs with no discontinuity in the density (see top right panel). However, this phase transition is a theory artifact since we know from the MC simulation that no singularity is apparent in the GEM4 at very high temperature, unless the pressure is really huge (cf. the case of $T=0.5$ discussed in the text).

\newpage
\section*{Figure captions}
%
%
{\bf FIGURE 1}. (Color online). Zero-temperature chemical potential of various 1D lattices relative to the ordinary (``1'') crystal, $\Delta\mu=\mu-\mu_1$, plotted as a function of the pressure $P$. Solid lines: ``2'', blue; ``3'', cyan; ``4'', magenta; ``5'', red; ``12'', green; ``23'', yellow (see main text and appendix A for notation). Dotted lines: ``112'', blue; ``122'', red. Dashed lines: ``1112'', blue; ``1122'', green; ``1222'', red. In the top right panel, the best ``1122'' crystal is a ``2'' crystal from $P=0.834$ to $P=2.465$, and a ``3'' crystal for larger $P$'s (eventually, for much higher pressures, the optimal ``1122'' becomes a ``6'' crystal). The vertical dotted lines mark the transition pressures (see Table 1). Note the abrupt change of slope of all $\Delta\mu$ curves at $P\simeq 1.46$ (left panel), which is due to a jump in the density of the stable ordinary crystal. On the other hand, the cusp seen in $\mu_2-\mu_1$ at $P\simeq 5.85$ reflects a density jump of the best 2-cluster crystal.

%
%
{\bf FIGURE 2}. (Color online). Plot of $\langle\Delta U\rangle_\lambda$ as a function of $\lambda$ for $n_c=2$ (for $T=0.01,\rho=1.5$, and $N=400$). The calculation was made using a reference system of $N$ independent particles moving in a 1D lattice of potential barriers~\cite{Mladek1} (barrier height: $10k_BT$; barrier width: $0.1$). The red straight-line segments which are drawn over the data points provide an independent estimate of the $\lambda$ derivative of $\langle\Delta U\rangle_\lambda$, extracted from the average square fluctuation of $\Delta U$. In the insets, I show that the chemical potentials computed at $T>0$ (open dots) are fully consistent with the respective $T=0$ (cluster-crystal) values (full dots). Left: $P=0.5$ path (``1'' phase). Right: $P=1.6$ path (``2'' phase).

%
%
{\bf FIGURE 3}. (Color online). Isothermal-isobaric MC simulation runs of the 1D GEM4 system: equilibrium number density $\rho$ vs. pressure $P$ along a number of isothermal paths. Each chain of runs refers to distinct initial values for $n_c$ {\em and} pressure: $n_c=1$ and $P=0.3$ or 0.4 (red full dots); $n_c=2$ and $P=2$ (black open squares); $n_c=3$ and $P=3$ or 4 (blue open dots); $n_c=4$ and $P=6$ (black open dots). In each panel, the thin solid lines represent the zero-temperature density curves for $n$-cluster crystals ($n=1,2,3,4$). The observed density jumps are associated with (metastable) $T=0$ isostructural transitions of the 1- and 2-cluster crystals (see also Fig.\,1 caption). The vertical dotted lines mark the transition pressures at $T=0$ (see Table I).

%
%
{\bf FIGURE 4}. (Color online). 1D GEM4 phase diagram on the $P$-$T$ plane (left) and on the $\rho$-$T$ plane (right). In the upper panels, I have reported some transition points between a standard fluid phase and a cluster-crystal phase of varying $n_c$, as computed through the DFT theory illustrated in appendix B (see also Fig.16). As argued in the text, this first-order line would actually be an artifact since it is not confirmed by simulation. The coexistence points reported in the lower panels were determined through the crossing of chemical-potential curves constructed by the method explained in the text. In the lower-right panel the extent of the coexistence regions (yellow shaded regions) can be appreciated. Here, the inverse rate of system compression/decompression was $2\times 10^6$ equilibrium sweeps per pressure (on isotherms) or temperature point (on isobars). If the runs were much longer a different estimate of the critical point had been obtained, see Sect.\,III. Farther from criticality the location of the transition points is insensitive to the length of the runs.

%
%
{\bf FIGURE 5}. (Color online). Left: $\langle\Delta U\rangle_\lambda$ plotted as a function of $\lambda$, for $T=0.5$ and $\rho=3.5$ ($n_c=2$, blue; 3, cyan; 4, green; 5, magenta; 6, red). The arrow marks the direction of increasing $n_c$. The small black straight-line segments over the data points for $n_c=5$ are MC estimates of the local slope of $\langle\Delta U\rangle_\lambda$, computed from the average square fluctuation of $\Delta U$ (see Fig.\,2). Right: Chemical potential for $T=0.5$, plotted as a function of pressure, for $n_c=2,3,4,5,6$ (top to bottom). For making the comparison easier, the chemical potential for $n_c=1$ has been subtracted from each curve (absolute $\mu_1$ values go from about 4 to about 6.5 in the $P$ range from 5 to 12).

%
%
{\bf FIGURE 6}. (Color online). 1D GEM4 system along the $T=0.5$ isotherm: energy and density data for a number of $n_c$ values ($n_c=1$, crosses; $n_c$ from 2 to 6, dots). The data points for $n_c=2,\ldots,6$ are hardly distinguishable one from another. The red solid line in the bottom panel is the MF-DFT estimate of the number density.

%
%
{\bf FIGURE 7}. (Color online). RDF of the 1D GEM4 system at $T=0.01$ and $P=1.6$ (2-cluster fluid state), plotted as a function of $r/\rho$. Top: linear-linear scale. Bottom: log-linear scale.

%
%
{\bf FIGURE 8}. (Color online). Number density of the 1D GEM4 system along three isotherms: $T=0.02$ (top), $T=0.016$ (center), and $T=0.01$ (bottom). The data points refer to a number of compression (dots) and decompression trajectories (squares). Different colors denote different numbers $M$ of sweeps produced per state point: $2\times 10^6$ (red), $5\times 10^6$ (magenta), $10^7$ (cyan), $2\times 10^7$ (blue), and $5\times 10^7$ (black). The arrows in the center and bottom panels mark the direction of increasing $M$. For each value of $T$, the data points for the most accurate sampling have been joined by straight-line segments. Plotted in green are the density values obtained from simulations of cluster crystals prepared with non-integer $n_c$ values: 1.8 (crosses) and 1.9 (diamonds). 

%
%
{\bf FIGURE 9}. (Color online). Isobaric specific heat (top) and isothermal compressibility (bottom) of the 1D GEM4 system along two isotherms: $T=0.016$ ($5\times 10^7$ sweeps per point, dashed blue lines) and $T=0.02$ ($2\times 10^7$ sweeps per point, solid red lines). Values of $c_P$ and $K_T$ along both compression (dots) and decompression trajectories (squares) are reported.

%
%
{\bf FIGURE 10}. (Color online). Pressure evolution of the mean site occupancy $n_c$, computed by Eq.\,(\ref{3-1}), along two isothermal paths at $T=0.02$ (top) and $T=0.016$ (bottom). Data from both the compression (blue dots) and the decompression trajectory (red squares) for the most accurate sampling are shown.

%
%
{\bf FIGURE 11}. (Color online). 1D GEM4 system along the $T=0.05$ isotherm. The data points refer to both compression (dots) and decompression trajectories (squares). Different colors denote different numbers $M$ of sweeps produced per state point: $2\times 10^6$ (red), $10^7$ (cyan), and $2\times 10^7$ (blue). The arrows in the top panel mark the direction of increasing $M$. The data points for the latter case are joined by straight-line segments. Top: number density. Bottom: mean site occupancy for the most accurate sampling (compression route, blue dots; decompression route, red squares).

%
%
{\bf FIGURE 12}. (Color online). Growth and coarsening kinetics of the 1D GEM4 system at $T=0.016$ and $P=0.89$. Top: plots of $G(k)$ (see text) for five different choices of the initial system configuration (from left to right, this configuration was extracted from the compression trajectory of Fig.\,8 center with $2,5,10,20,50$ million equilibrium sweeps per state point, respectively). The different colors denote MC time, increasing in steps of $5\times 10^7$ sweeps from red to black (i.e., roughly from left to right within each panel). Bottom: fraction $f_2$ of lattice sites occupied by particle pairs, plotted as a function of MC time. The different colors now denote the number of million sweeps generated per state point in the compression trajectory from which the initial configuration of the present run was extracted: 2 (red), 5 (magenta), 10 (cyan), 20 (blue), and 50 (black), increasing from bottom to top at small $t$.

%
%
{\bf FIGURE 13}. (Color online). Same as in Fig.\,12, but now for $T=0.01$ and $P=1.05$. Top panels refer to four different choices of the initial configuration, which were now extracted from the compression trajectories seen in the bottom panel of Fig.\,8 (from left to right, the inverse compression rate is $2,5,10,20$ million sweeps, respectively). Bottom: time evolution of $f_2$ for the same four initial configurations.

%
%
{\bf FIGURE 14}. (Color online). Final system configuration of the $3\times 10^8$-sweeps-long runs that were produced at $T=0.016,P=0.89$ (top panel) and $T=0.01,P=1.05$ (bottom panel) in order to investigate the growth dynamics of ``2'' from ``1''. The blue (i.e., darker) dots are isolated particles, whereas each cyan dot marks a particle which is tightly bound to another particle, thus forming a pair with it. The total number of particles here is $N=400$, and the particle coordinate increases from left to right and from top to bottom. The numbers below each panel of eight $40\sigma$-long slices are abscissas in $\sigma$ units for the last slice (note that periodic boundary conditions are applied at the box boundaries, the box length being $V/\sigma=297.0585$ for $T=0.016$ and $V/\sigma=280.3364$ for $T=0.01$).

%
%
{\bf FIGURE 15}. (Color online). DFT in the mean-field approximation for the 3D GEM4 system. The data are for a number of temperatures (from left to right, $T=0.1,0.3,0.5,1$) and are plotted as a function of the reference-fluid density $\rho$. For each $T$, the two vertical dotted lines mark the phase transitions (the leftmost is from fluid to bcc cluster crystal, the rightmost is from bcc to fcc cluster crystal). Top left panel: grand-potential difference between solid and fluid (fcc, blue solid line; bcc, red dashed line). Top right panel: solid density. Bottom left panel: $\alpha$ parameter. Bottom right panel: mean site occupancy.

%
%
{\bf FIGURE 16}. MF-DFT results for the 3D GEM4 system. Top: $P$-$T$ phase diagram. Bottom: $\rho$-$T$ phase diagram. The width of the (cluster bcc)-(cluster fcc) coexistence region is negligible on the scale of the figure.

%
%
{\bf FIGURE 17}. (Color online). MF-DFT results for the 1D GEM4 system. Data for a number of temperatures are shown (from left to right, $T=0.3,0.5,0.7,1$) and plotted as a function of the reference-fluid density $\rho$. For each $T$, the vertical dotted line marks the phase transition from fluid to cluster crystal. Top left panel: grand-potential difference between solid and fluid. Top right panel: solid density (the thin red straight line is $\rho\sigma$). Bottom left panel: $\alpha$ parameter. Bottom right panel: mean site occupancy.

\newpage
%
%
\begin{figure}
\centering
\includegraphics[width=16cm]{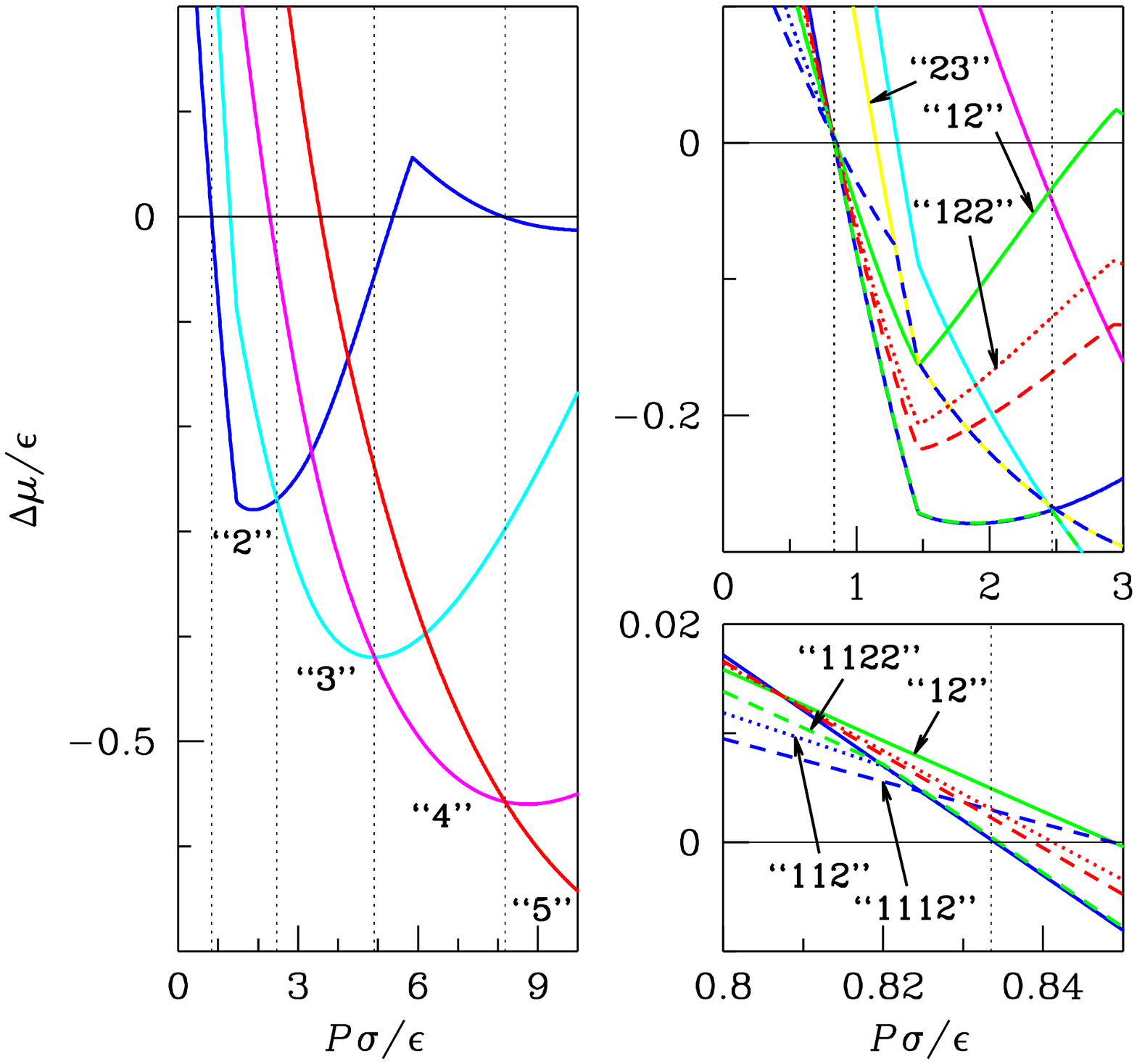}
\caption{
}
\label{fig1}
\end{figure}

%
%
\begin{figure}
\centering
\includegraphics[width=16cm]{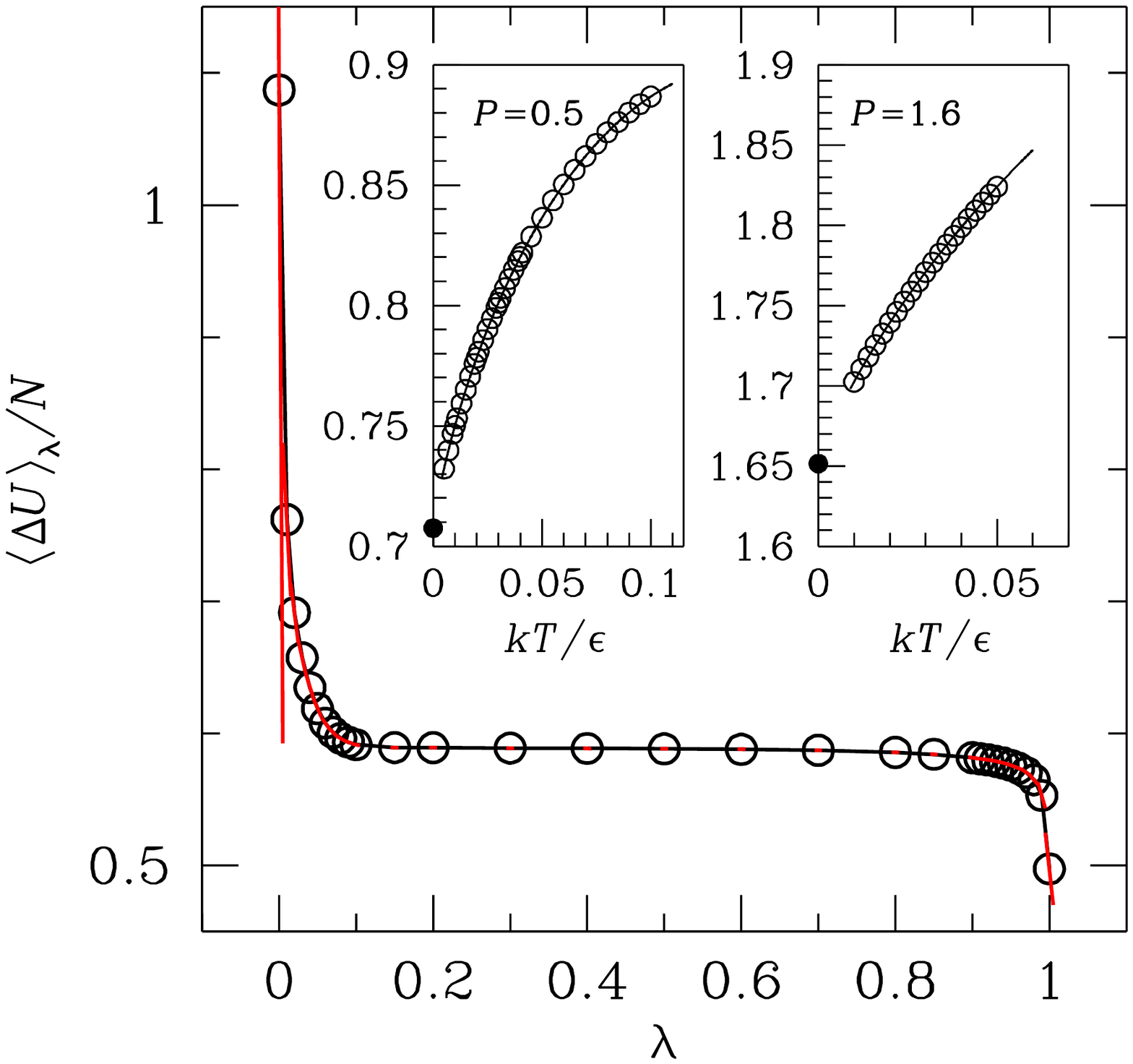}
\caption{
}
\label{fig2}
\end{figure}

%
%
\begin{figure}
\centering
\includegraphics[width=16cm]{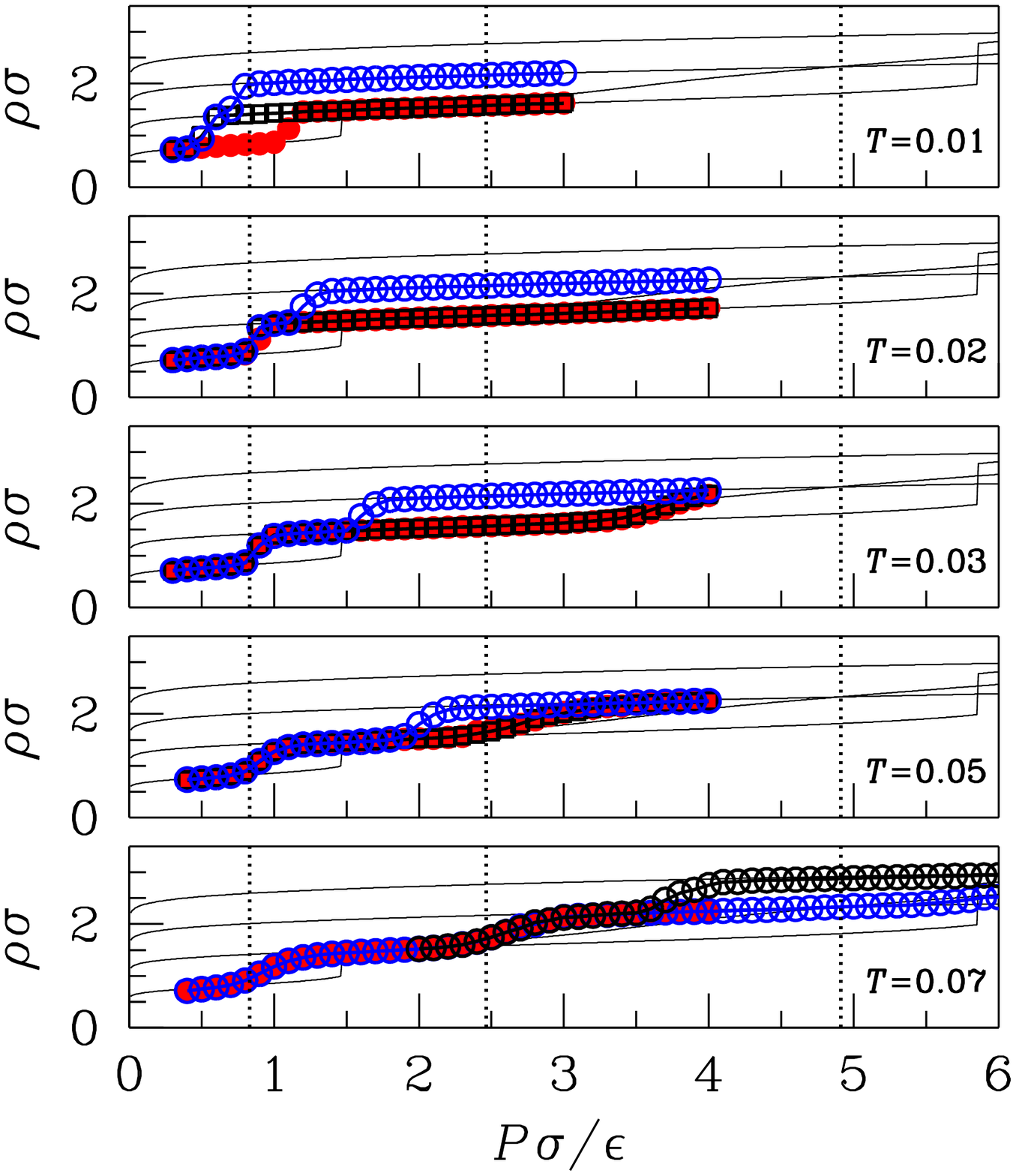}
\caption{
}
\label{fig3}
\end{figure}

%
%
\begin{figure}
\centering
\includegraphics[width=16cm]{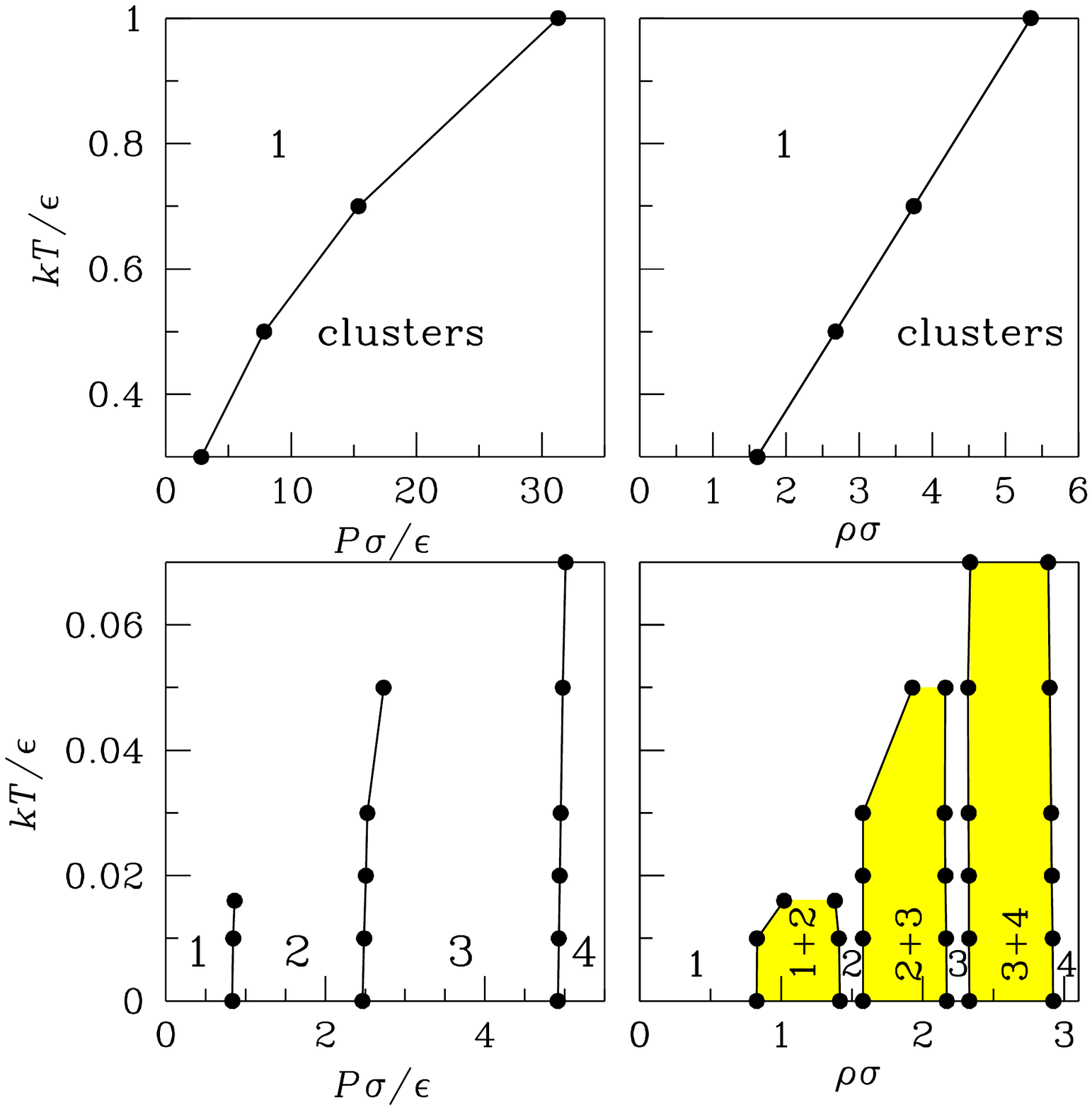}
\caption{
}
\label{fig4}
\end{figure}

%
%
\begin{figure}
\centering
\includegraphics[width=16cm]{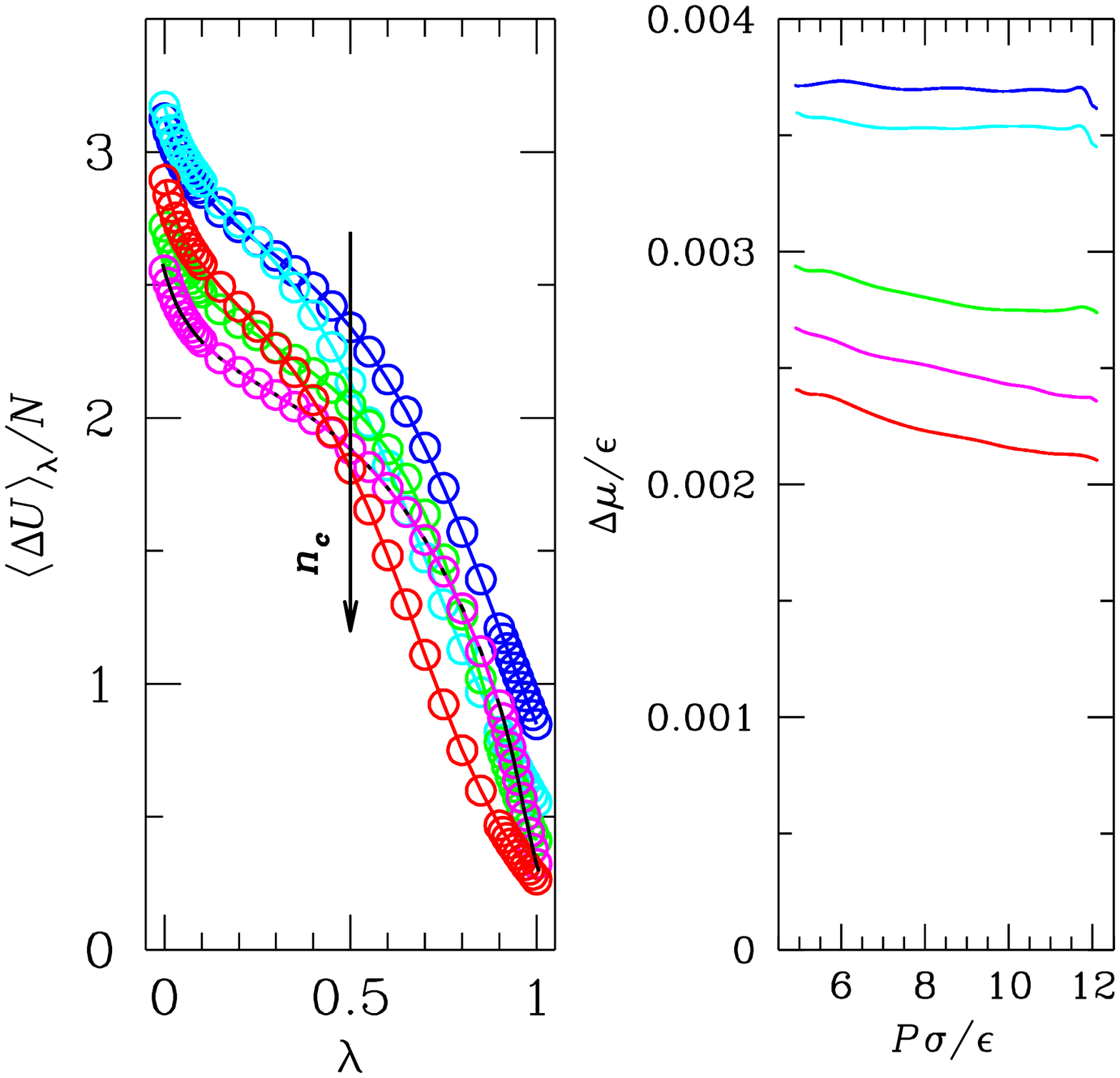}
\caption{
}
\label{fig5}
\end{figure}

%
%
\begin{figure}
\centering
\includegraphics[width=16cm]{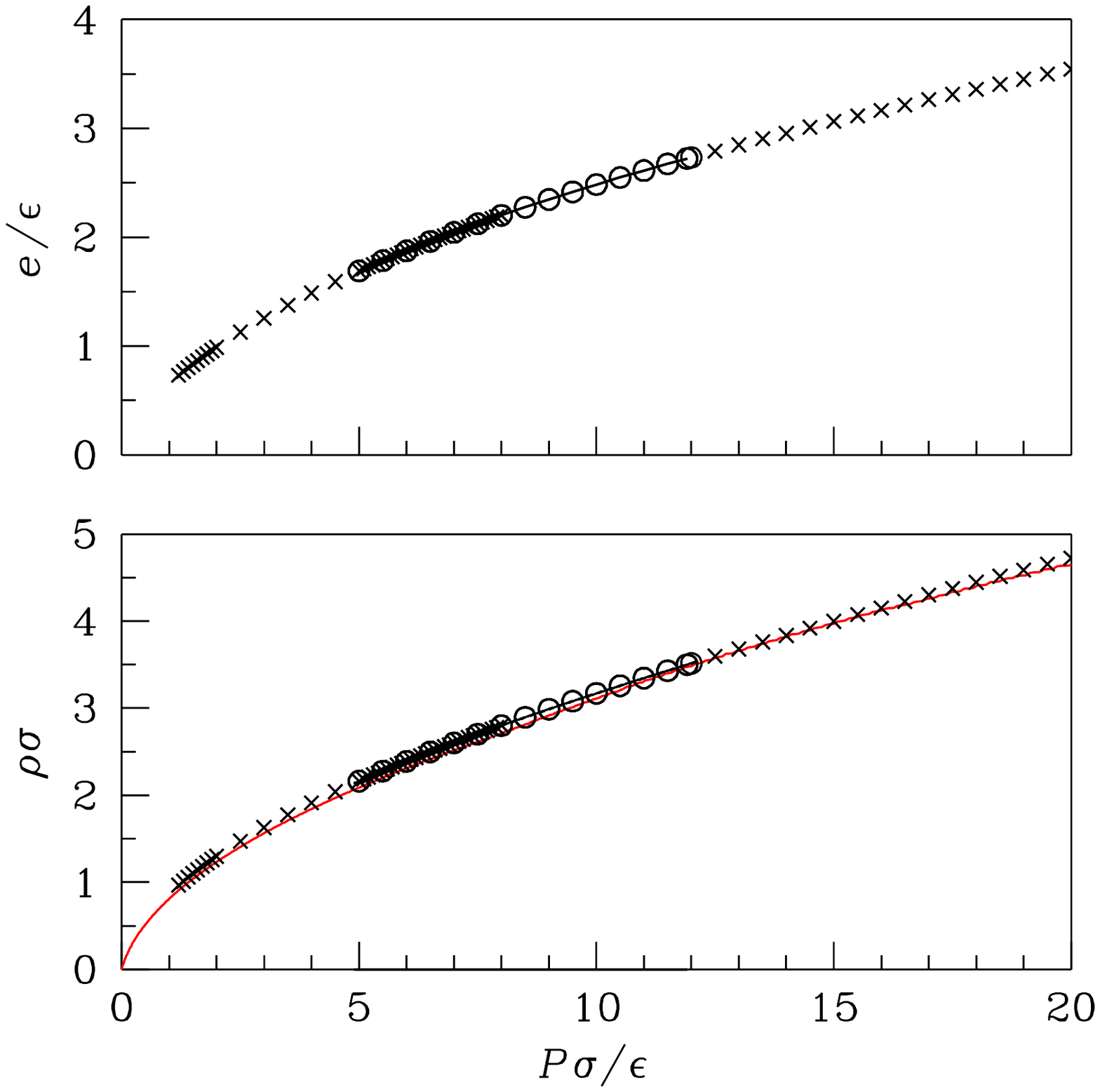}
\caption{
}
\label{fig6}
\end{figure}

%
%
\begin{figure}
\centering
\includegraphics[width=16cm]{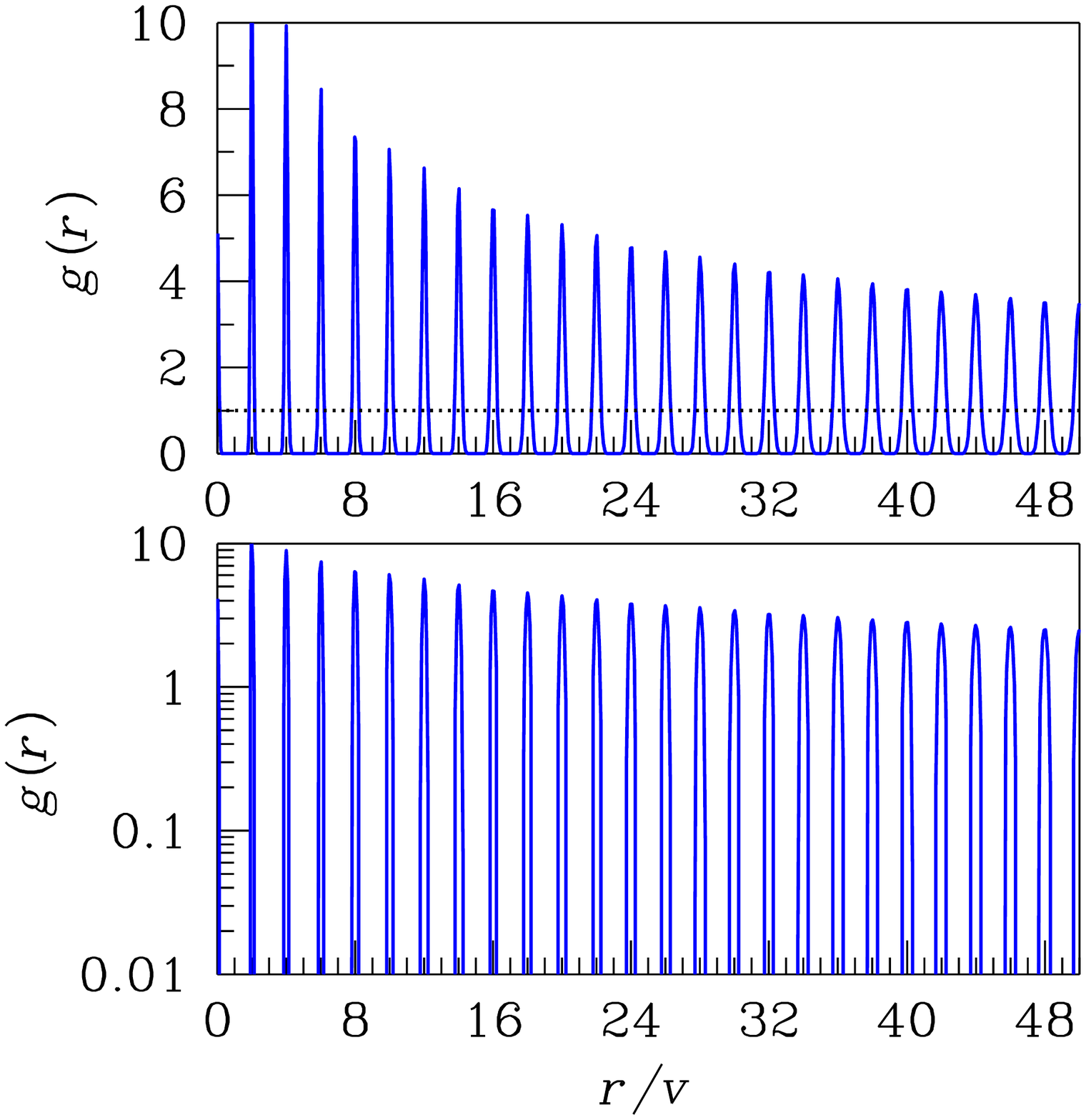}
\caption{
}
\label{fig7}
\end{figure}

%
%
\begin{figure}
\centering
\includegraphics[width=16cm]{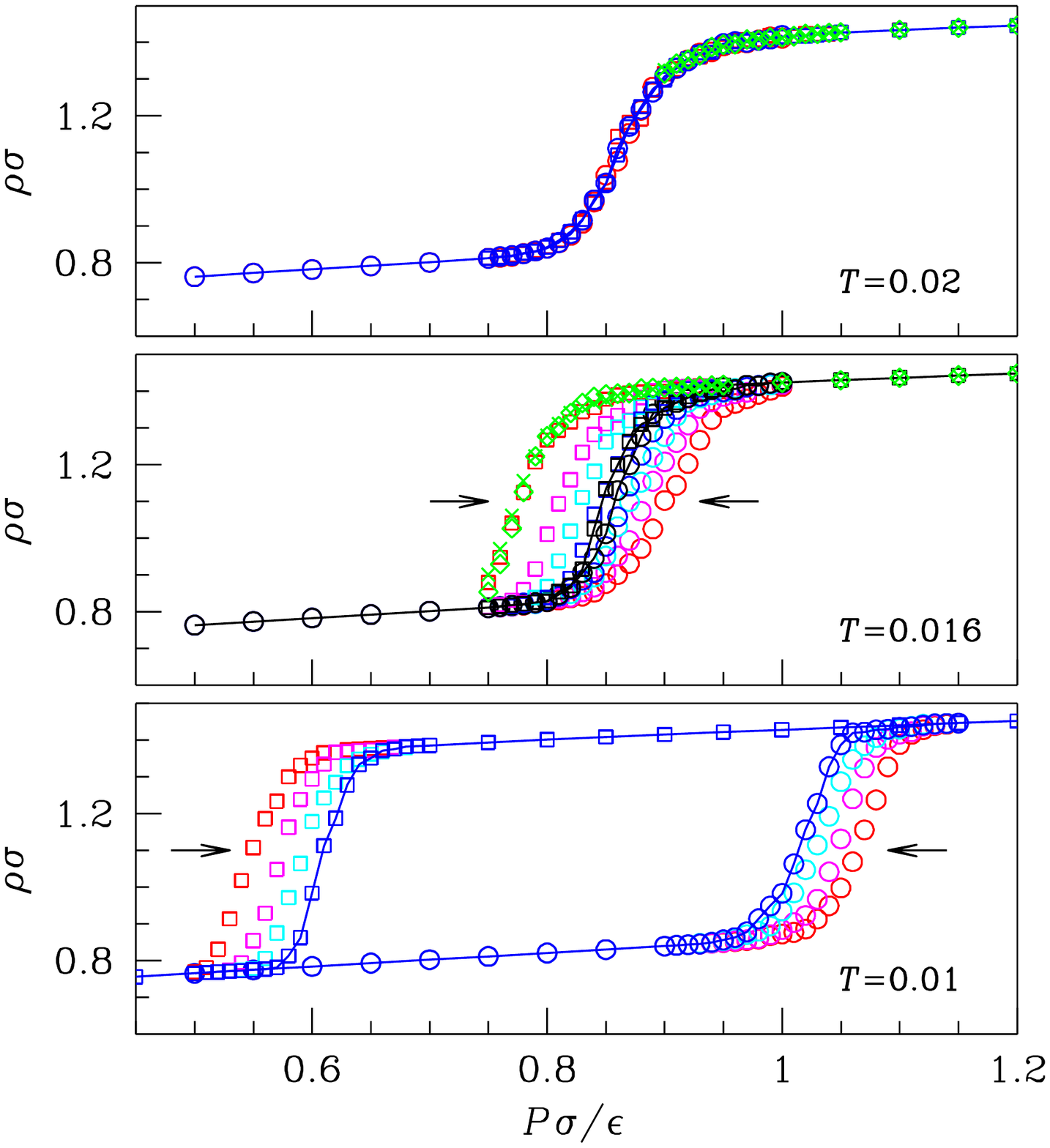}
\caption{
}
\label{fig8}
\end{figure}

%
%
\begin{figure}
\centering
\includegraphics[width=16cm]{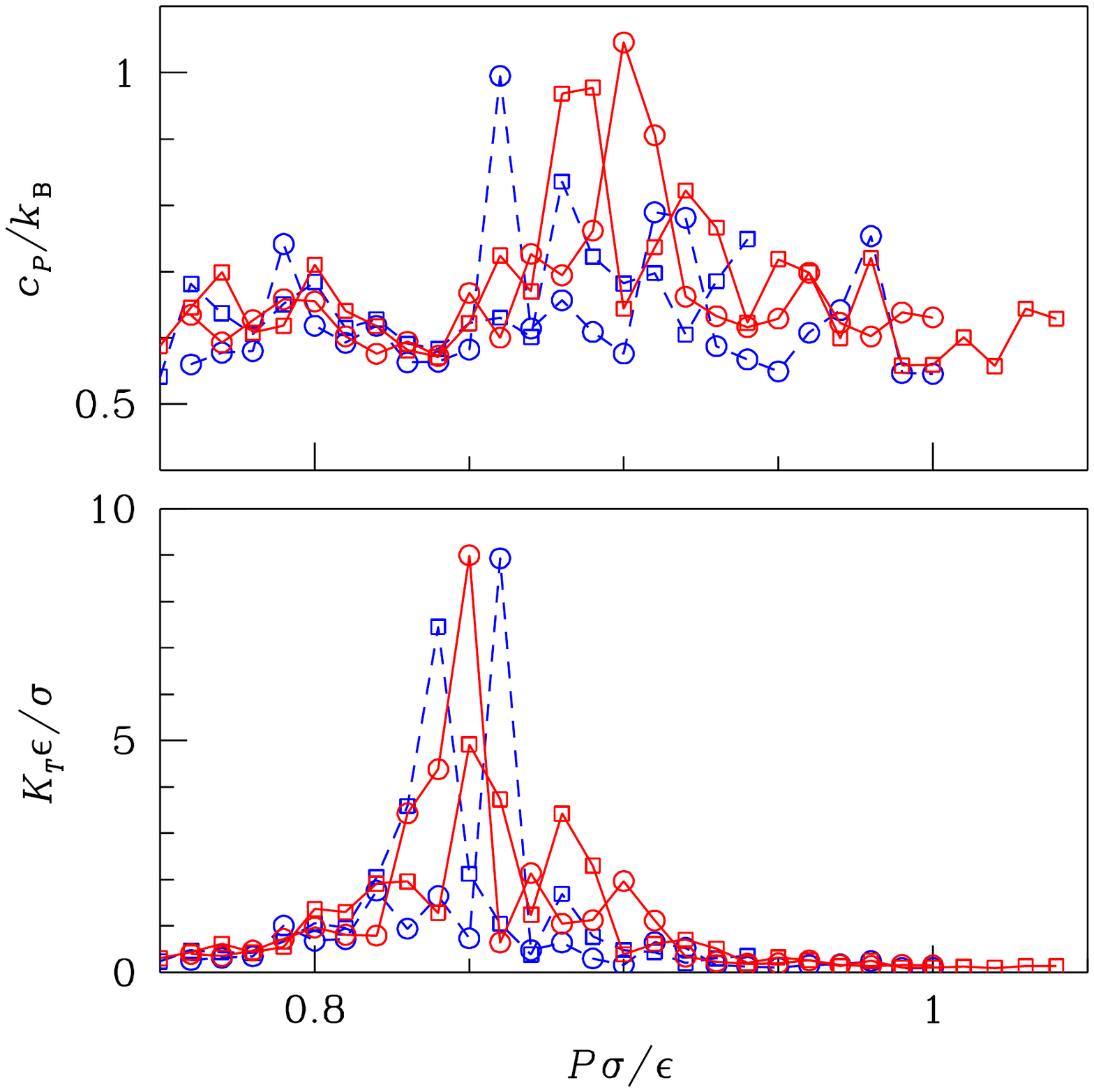}
\caption{
}
\label{fig9}
\end{figure}

%
%
\begin{figure}
\centering
\includegraphics[width=16cm]{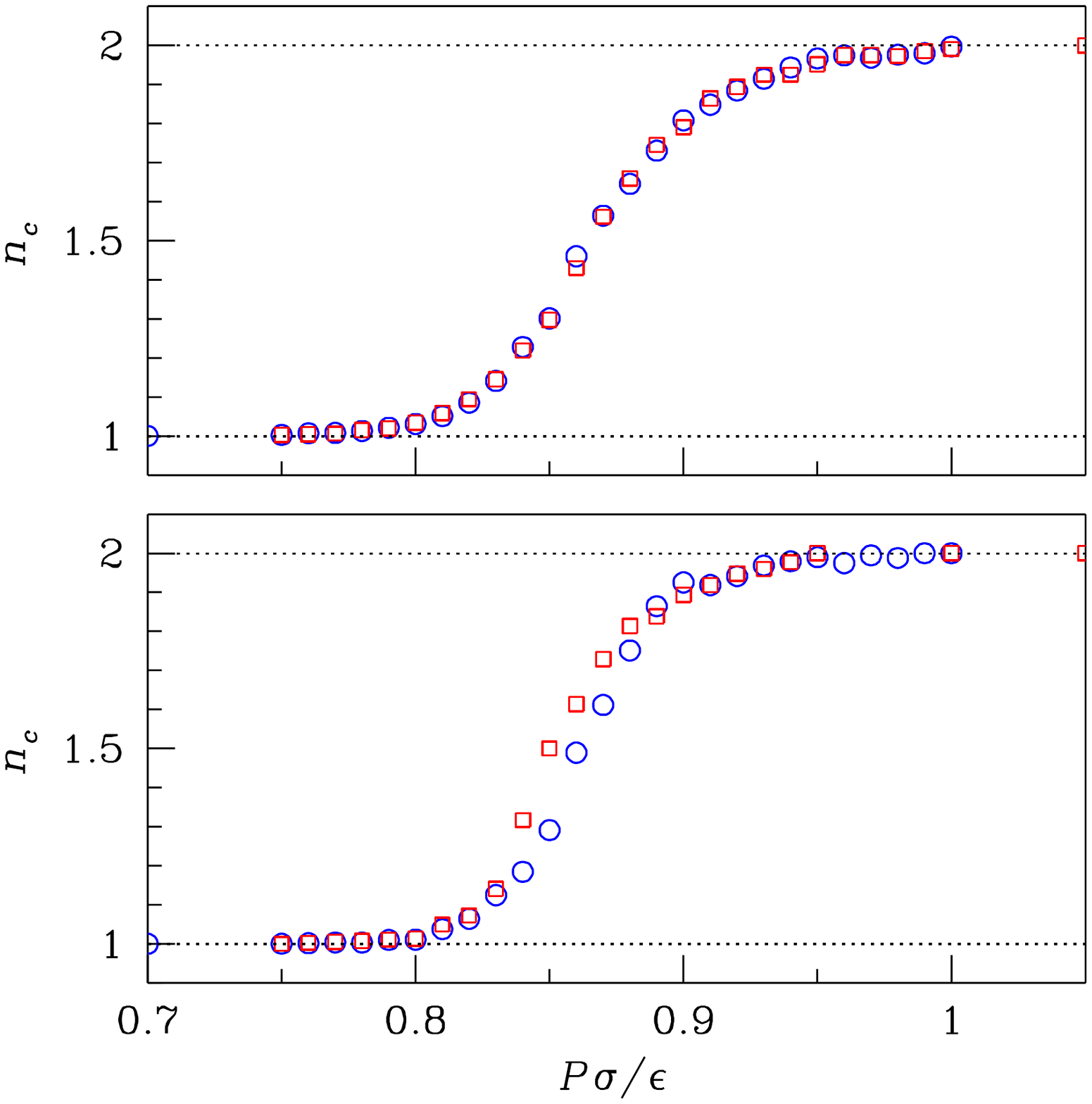}
\caption{
}
\label{fig10}
\end{figure}

%
%
\begin{figure}
\centering
\includegraphics[width=16cm]{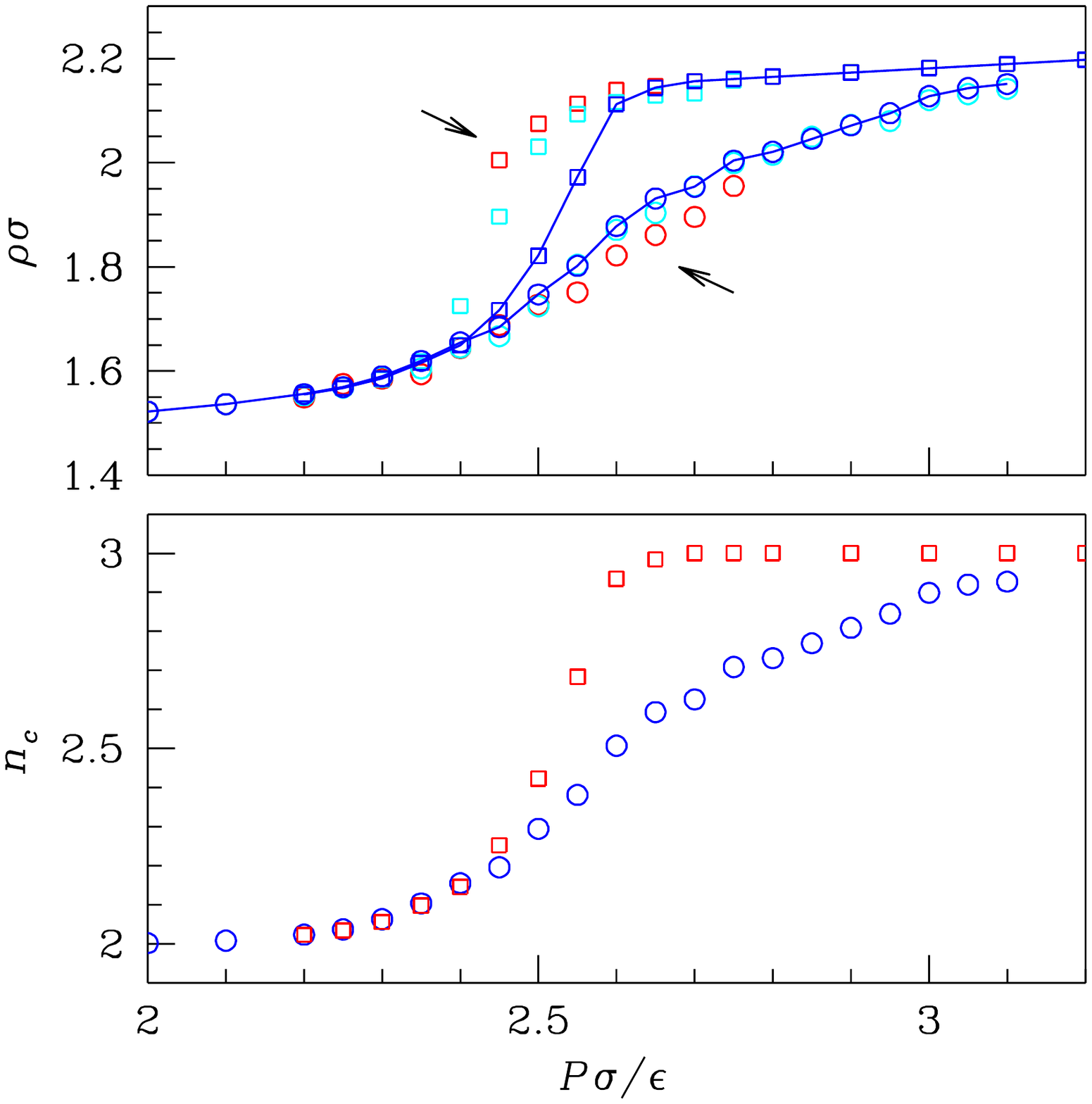}
\caption{
}
\label{fig11}
\end{figure}

%
%
\begin{figure}
\centering
\includegraphics[width=16cm]{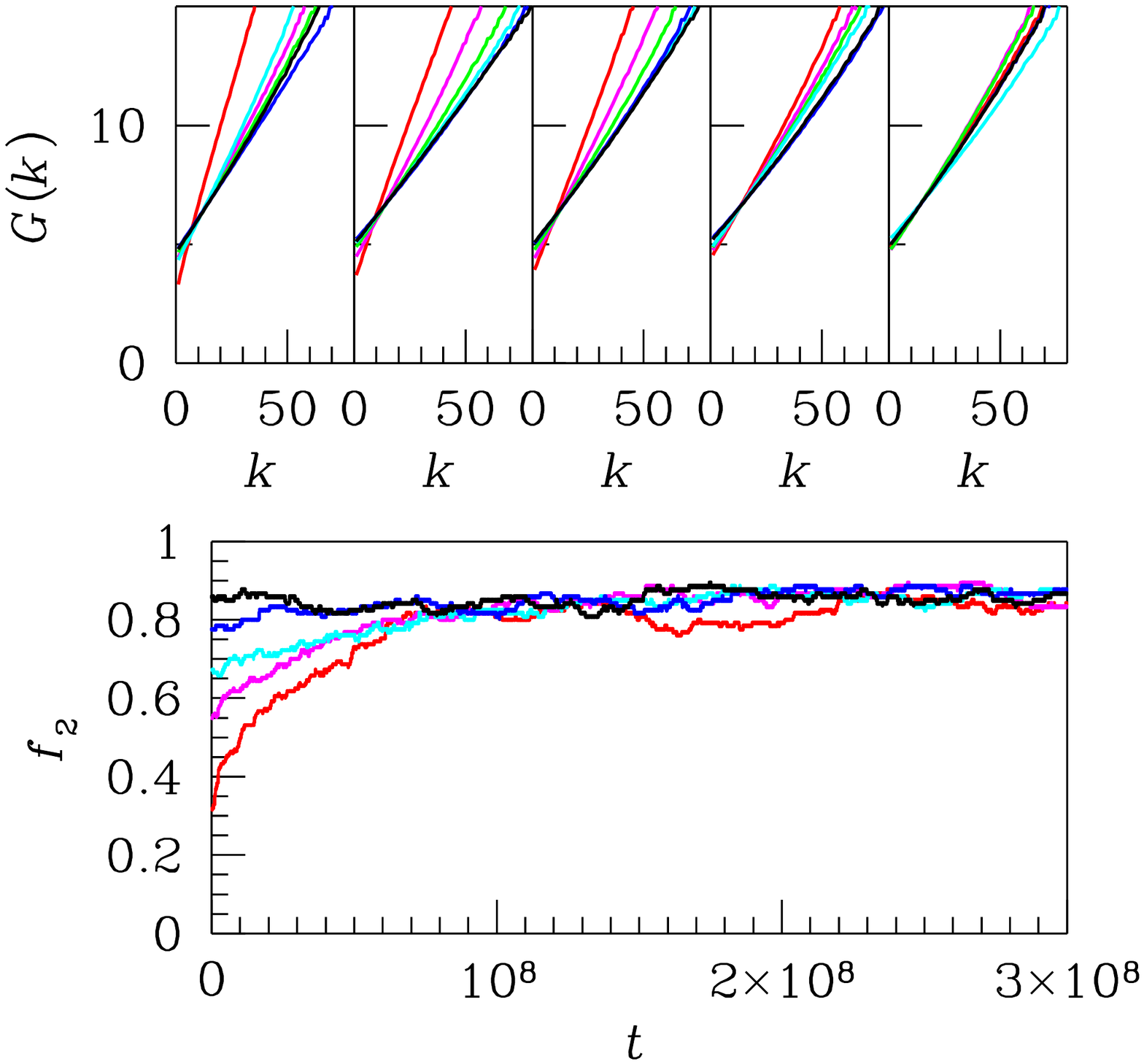}
\caption{
}
\label{fig12}
\end{figure}

%
%
\begin{figure}
\centering
\includegraphics[width=16cm]{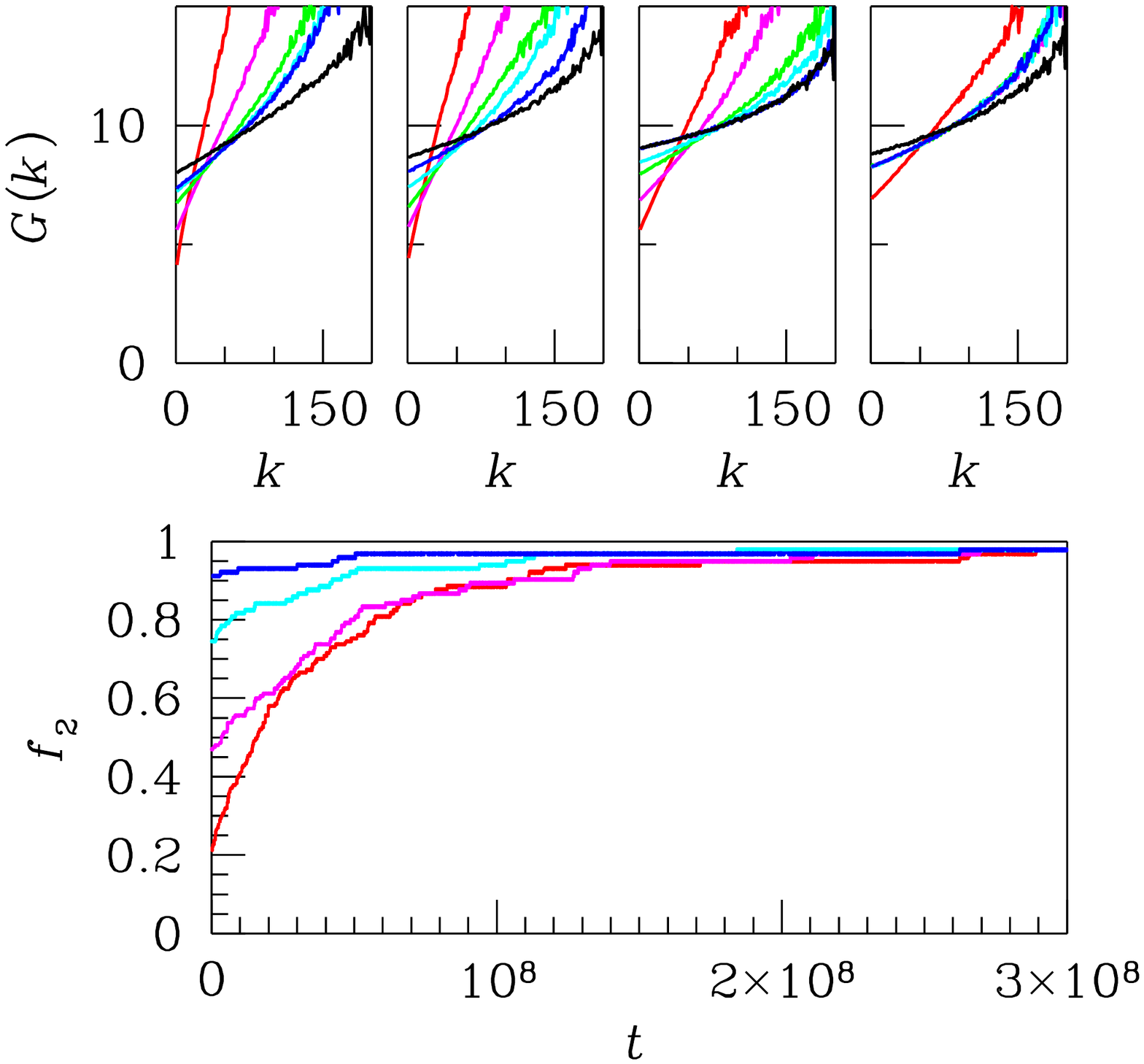}
\caption{
}
\label{fig13}
\end{figure}

%
%
\begin{figure}
\centering
\includegraphics[width=16cm]{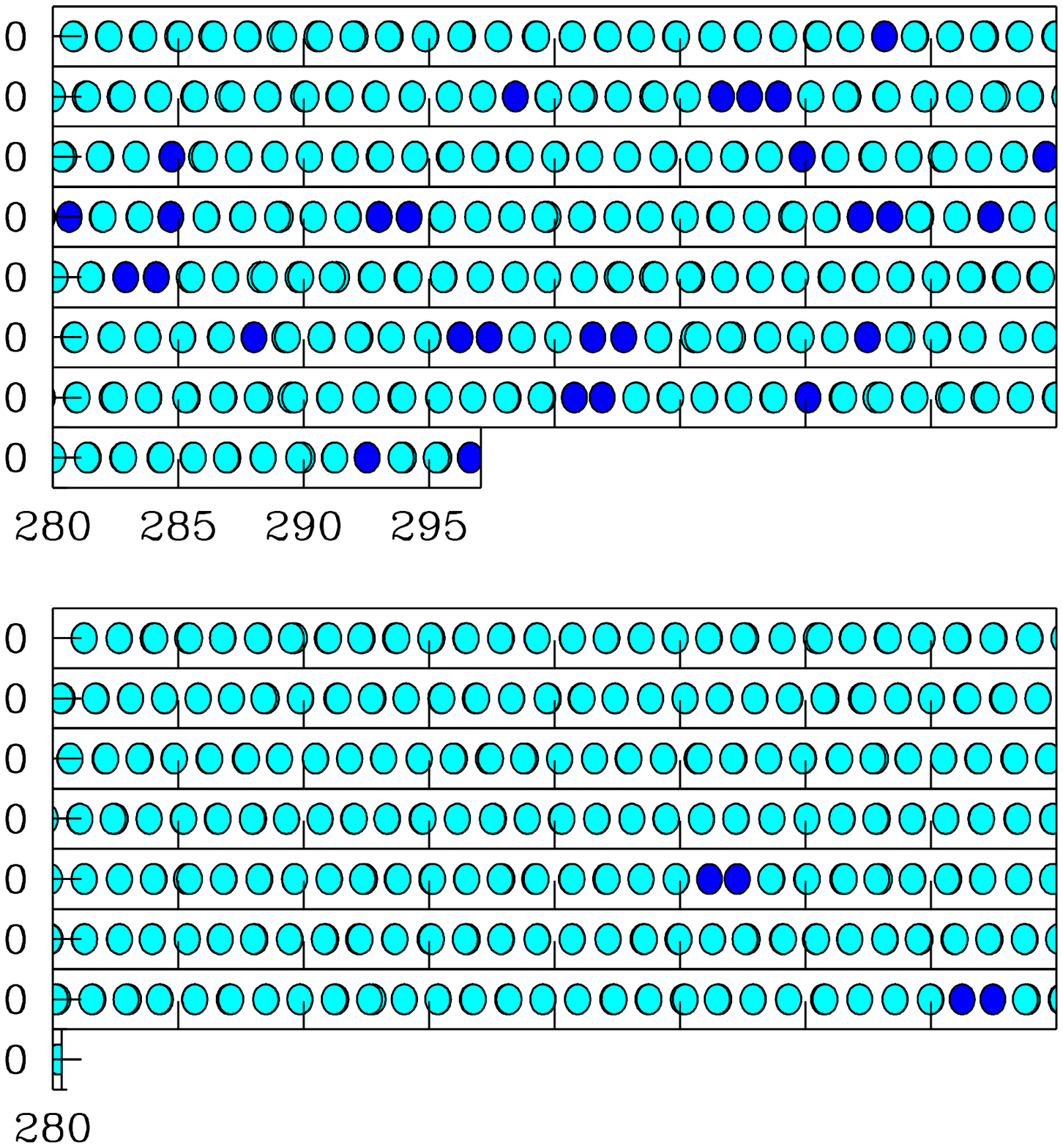}
\caption{
}
\label{fig14}
\end{figure}

%
%
\begin{figure}
\centering
\includegraphics[width=16cm]{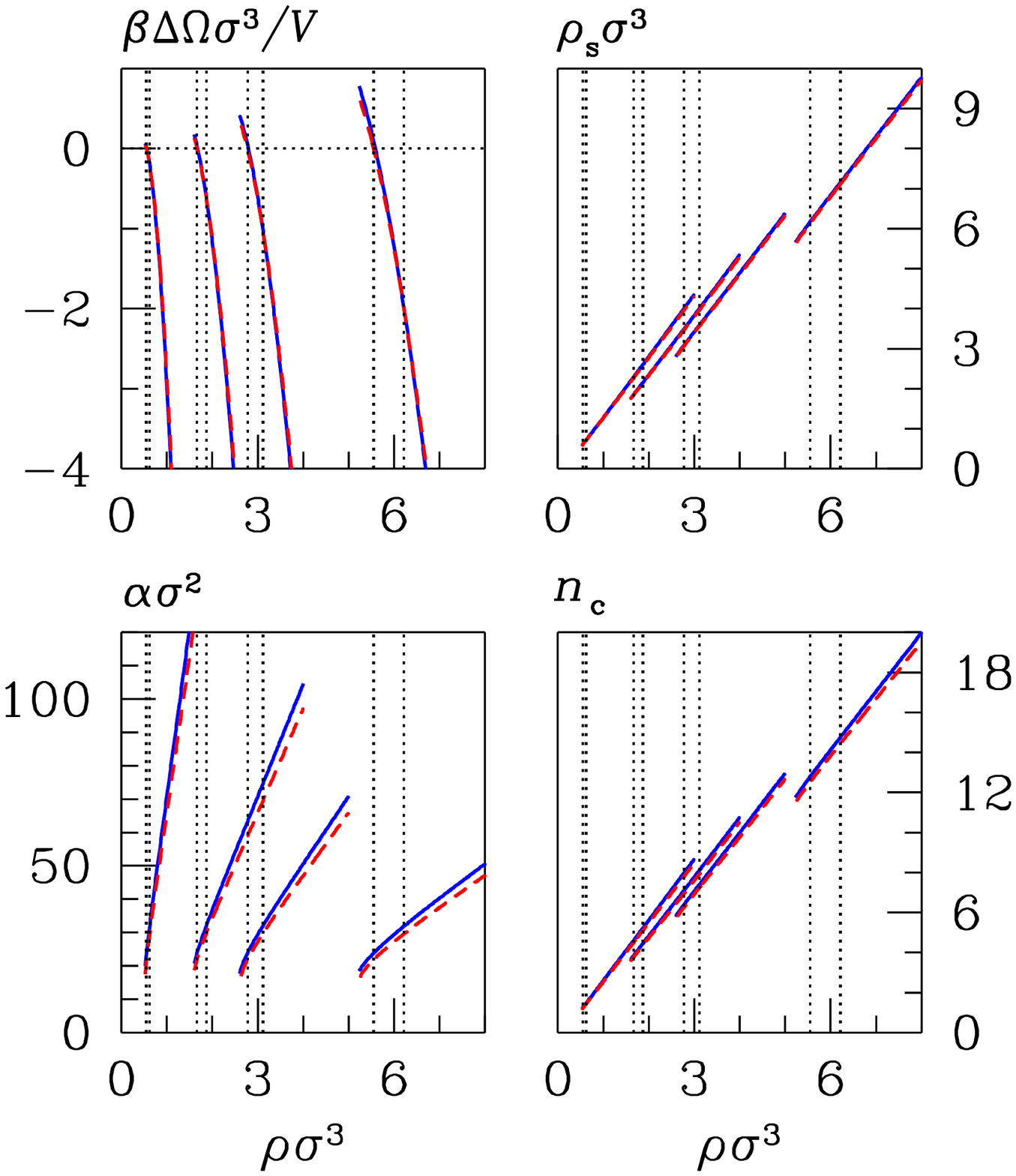}
\caption{
}
\label{fig15}
\end{figure}

%
%
\begin{figure}
\centering
\includegraphics[width=16cm]{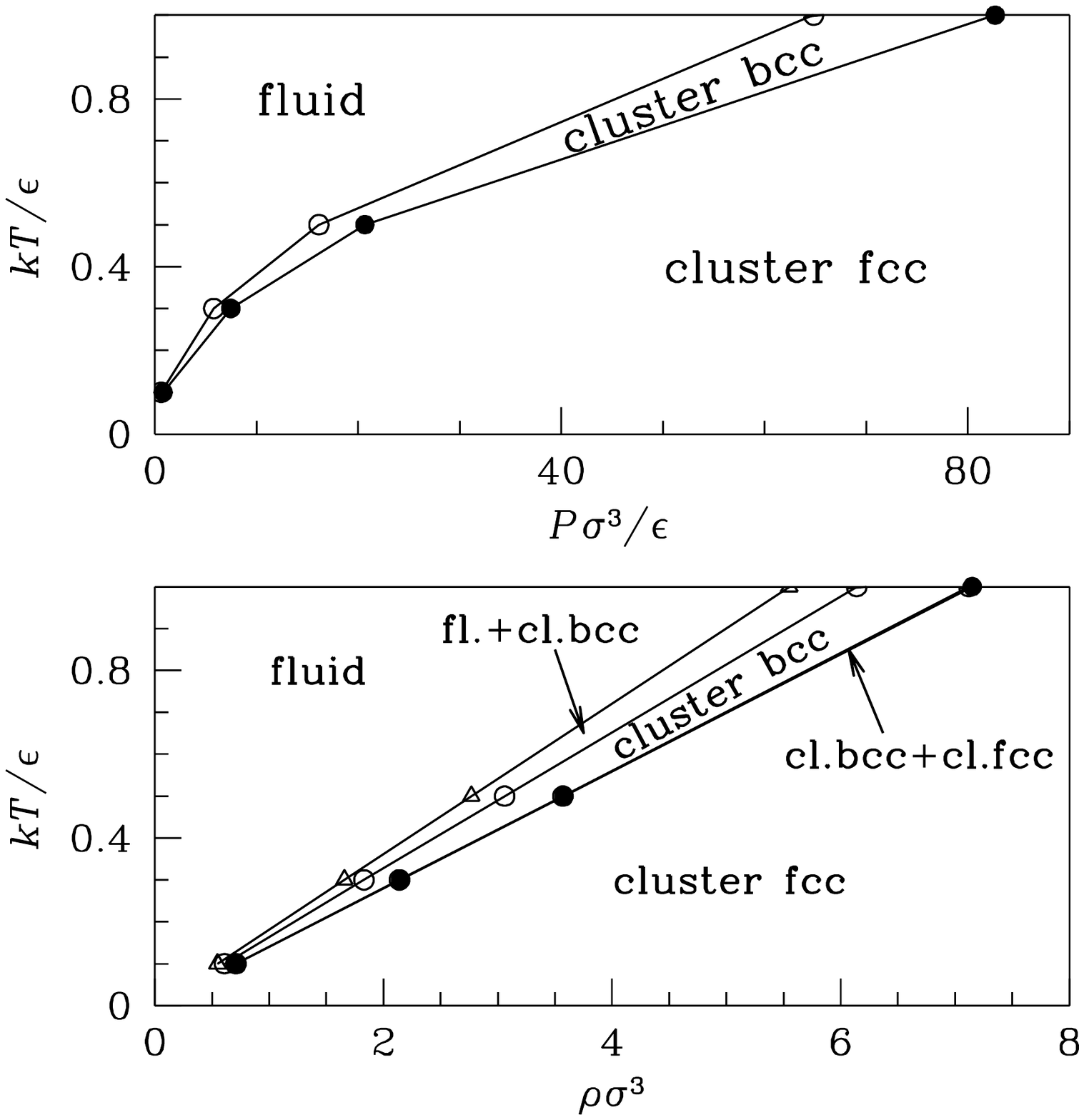}
\caption{
}
\label{fig16}
\end{figure}

%
%
\begin{figure}
\centering
\includegraphics[width=16cm]{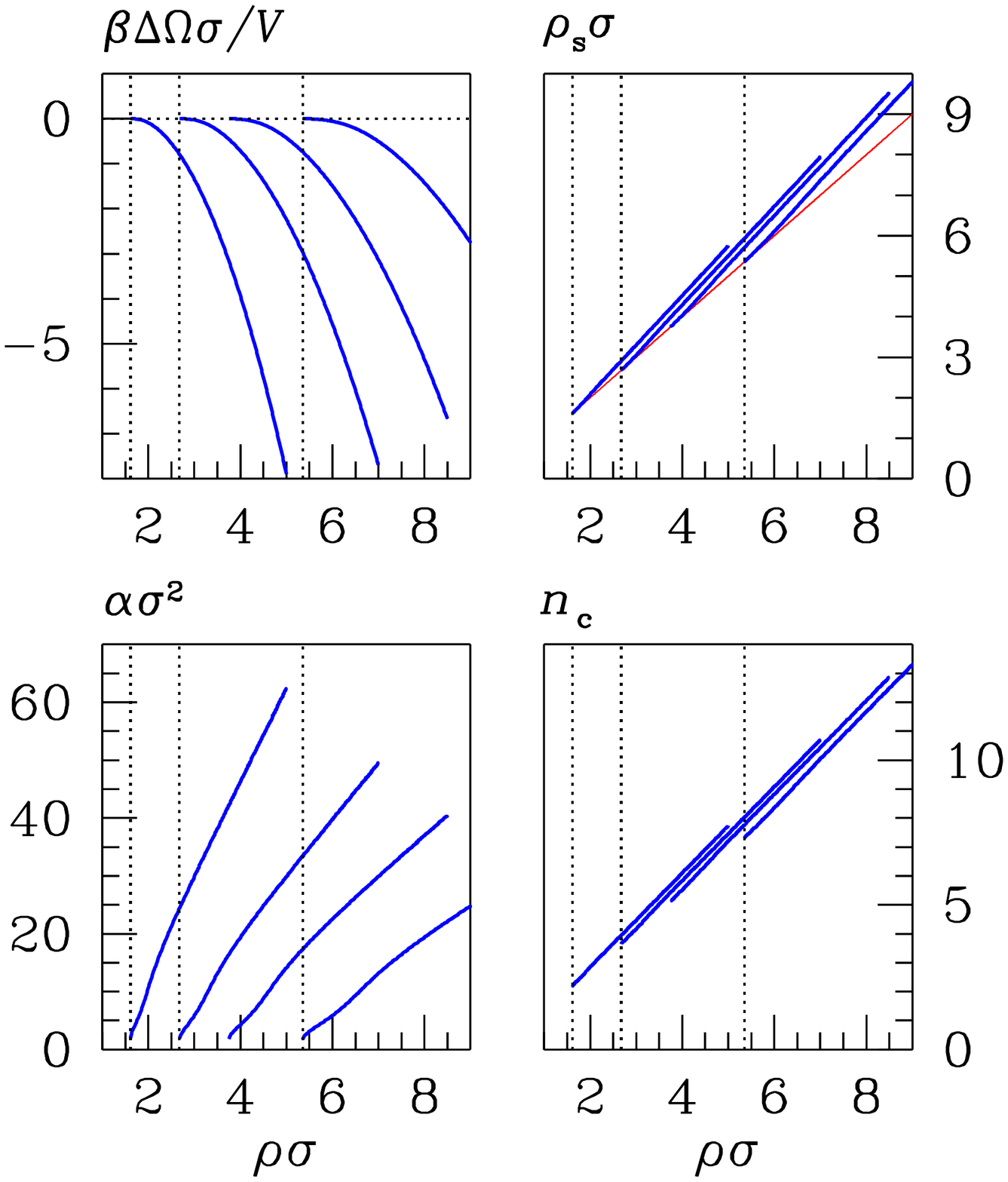}
\caption{
}
\label{fig17}
\end{figure}
\end{document}